\newcommand{\diracslash}[1]{#1\llap{/\kern2pt}}

\newcommand{\be}{\begin{equation}}
\newcommand{\ee}{\end{equation}}
\newcommand{\bea}{\begin{eqnarray}}
\newcommand{\eea}{\end{eqnarray}}
\newcommand{\ba}[1]{\begin{array}{#1}}
\newcommand{\ea}{\end{array}}

\newcommand{\bt}{\begin{tabular}}
\newcommand{\et}{\end{tabular}}

\newcommand{\beas}{\begin{eqnarray*}}
\newcommand{\eeas}{\end{eqnarray*}}

\documentclass[preprint,prd,aps,floats,nofootinbib,floatfix]{revtex4}

\DeclareSymbolFont{rsfs}{U}{rsfs}{m}{n}
\DeclareSymbolFontAlphabet{\mathrsfs}{rsfs}
\usepackage{graphicx}
\usepackage{multirow}

\usepackage{graphicx}
\usepackage{cleveref}

\begin{document}

\title{Heavy Vector and Axial-Vector Mesons in  Hot and Dense 
Asymmetric Strange Hadronic  Matter} 
%%%%%%%%%%%%%%%%%%%%%%%%%%%%%%%%%%%%%%%%%%%%%%%%%%%%%%%%%%%%%%%%%
\author{Arvind Kumar}
\email{iitd.arvind@gmail.com, kumara@nitj.ac.in}
\affiliation{Department of Physics, Dr. B R Ambedkar National Institute of Technology Jalandhar, 
 Jalandhar -- 144011,Punjab, India}
 \author{Rahul Chhabra}
\email{rahulchhabra@ymail.com,}
\affiliation{Department of Physics, Dr. B R Ambedkar National Institute of Technology Jalandhar, 
 Jalandhar -- 144011,Punjab, India}

\def\be{\begin{equation}}
\def\ee{\end{equation}}
\def\bearr{\begin{eqnarray}}
\def\eearr{\end{eqnarray}}
\def\zbf#1{{\bf {#1}}}
\def\bfm#1{\mbox{\boldmath $#1$}}
\def\hf{\frac{1}{2}}
\def\kp{\zbf k+\frac{\zbf q}{2}}
\def\km{-\zbf k+\frac{\zbf q}{2}}
\def\hwo{\hat\omega_1}
\def\hwt{\hat\omega_2}

\begin{abstract}

We  calculate the effects of finite density and temperature  of 
isospin asymmetric strange
hadronic matter, for different strangeness fractions,
on the in-medium properties of 
vector  $\left( D^{\ast}, D_{s}^{\ast}, B^{\ast}, B_{s}^{\ast}\right)$ 
and axial-vector  $\left( D_{1}, D_{1s}, B_{1}, B_{1s}\right)$ mesons,
using chiral hadronic SU(3) model and QCD sum rules.
We focus on the evaluation of in-medium mass-shift and shift in decay constant
 of above  vector and axial-vector mesons.
In QCD sum rule approach, the properties, e.g., the masses and decay constants
of vector and axial-vector mesons
are written in terms of quark and gluon condensates.
These quark and gluon condensates  are evaluated in the
present work within chiral SU(3) model,
through the medium modification of, scalar-isoscalar fields
$\sigma$ and $\zeta$, the scalar-isovector field
 $\delta$ and scalar dilaton field $\chi$, 
in the  strange hadronic medium which 
includes both nucleons as well as hyperons.
As we shall see in detail, the masses and decay constants
of heavy vector and axial-vector mesons 
are affected significantly due to isospin asymmetry 
and  strangeness fraction of the medium
and these modifications may influence
the experimental observables
produced in heavy ion collision experiments. 
The results of present investigations of in-medium properties of
vector and axial-vector mesons at
finite density  and temperature of strange hadronic medium may be helpful
for understanding the experimental data from heavy-ion collision experiments
in-particular for the 
Compressed Baryonic Matter (CBM) experiment of FAIR facility at
GSI, Germany. 


\end{abstract}

\maketitle
\section{Introduction}
The aim of 
relativistic heavy-ion collision experiments
is to explore the different phases of QCD phase diagram
so as to understand the underlying strong
interaction physics of
Quantum Chromodynamics. The different regions of QCD phase
diagram can be explored by varying the beam energy in the
high energy heavy-ion collision experiments.
The nucleus-nucleus collisions at the Relativistic Heavy Ion Collider (RHIC)
and Large Hadron Collider (LHC)
experiments explore the region of the QCD phase
diagram at low baryonic densities and high temperatures.
However, the objective of the Compressed Baryonic Matter (CBM)
experiment of FAIR project (at GSI Germany) is to study the region of phase diagram at high baryonic density and
moderate temperature. In nature these kind of phases may
exist in astrophysical compact objects, e.g., in neutron
stars. Among the many different observables  which may be produced
in CBM experiment, the one of them may be the production of mesons having
charm quark or antiquark.
Experimentally, charm meson spectroscopy as well as their
in-medium properties are also of interest from
the point of view of PANDA experiment of FAIR project,
where $\bar{p}A$ collisions will be performed.
The possibility of the production of open or hidden charm mesons
motivate the theoretical
physicist to study the properties of these mesons in
dense nuclear matter. The discovery
of many open or hidden charm
or bottom  mesons at CLEO, Belle or BABAR experiments \cite{cleo,bele,babar}
attract the attentions of theoretical groups to study the
properties of these mesons.

 In the present paper, our objective is to work out the
 in-medium masses and decay constants 
 of heavy charmed vector  
 $\left( D^{\ast}, D_{s}^{\ast} \right)$ 
and axial-vector  $\left( D_{1}, D_{1s}\right)$ 
as well as bottom vector $\left(B^{\ast}, B_{s}^{\ast}\right)$
and axial-vector $\left( B_{1}, B_{1s}\right)$ mesons
 in the isospin asymmetric strange hadronic medium, 
 at finite density and temperature
 The initial interest in the understanding of the properties of 
 heavy flavor mesons was stimulated from the
 observation of $J/\psi$ suppression phenomenon 
 \cite{sps,rhic1,alice2,rhic2}.
 In the high temperature regime, the $J/\psi$
 suppression can be explained either using the color screening 
 or statistical hadronization model \cite{satz,stat1,stat2,stat3,stat4}.
 However, for the moderate temperature and finite baryonic
 densities the effects of hadronic absorptions and comover interactions
 also come into picture and cannot be neglected \cite{cap1}.
The evaluation of in-medium properties of
open charm $D$ mesons may also play important role
in understanding of $J/\psi$ suppression
 in heavy-ion collisions.
 The reason is, the higher charmonium states are considered
as major source of $J/\psi$  mesons. However, if the $D$ mesons 
undergo mass drop in the nuclear matter and the in-medium
mass of $D\bar{D}$ pairs falls below the threshold value of 
excited charmonium states, then these charmonium states can also
decay to $D\bar{D}$ pairs and may cause a decrease in the yield of
$J/\psi$ mesons.  
The in-medium properties of
$D$ and $\bar{D}$ mesons may 
also play an important role in revealing the possibility of formation of charmed mesic nuclei. The charmed mesic nuclei are
the bound states of charm mesons and nucleon formed through strong interactions.
In Ref. \citep{qmc1}, the mean field potentials of $D$ and $\bar{D}$ mesons
were calculated using the quark meson coupling (QMC) model 
under local density approximation and the possibilities of the
formation of  bound states of $D^{-}$, $D^{0}$  and $\bar{D^{0}}$
mesons with Pb(208) were examined. The properties of charmed mesons in the nuclei
had also been studied using the unitary meson-baryon coupled
channel approach and incorporating the heavy-quark spin symmetry \citep{tolosbnd,tolosint91}.  
 The theoretical investigations of open or hidden charmed meson properties
at finite density and temperature of the nuclear matter
may help us in understanding their 
production rates, decay constants, decay widths, etc., in
heavy-ion collision experiments.

The in-medium properties of open charm and bottom mesons
are evaluated using the
 quark meson coupling model in Refs. \cite{qmc1,qmc2}.
The medium modified masses of $D$ mesons, calculated within QMC model,
were used as input for the
study of mass modification of $J/\psi$
mesons in the nuclear matter \cite{krein1}.
Borel transformed QCD sum rules,
involving operator product expansion (OPE) upto dimension-$4$,
 were used 
to investigate the in-medium modifications of
pseudoscalar $D$ mesons  
and the impact of these modifications on charmonium suppression
in Ref. \cite{haya1}.
The study of open charm mesons within QCD sum rules
were further carried out in Ref. \cite{higler1}, where
even as well as odd part of OPE
was used to study mass-spitting between pseudoscalar $D$-$\bar{D}$ doublet.
In the study of mass shift of $D$ mesons in Ref. \cite{higler1}, a positive mass shift of about $+45$ MeV 
was obtained at normal nuclear saturation density, whereas
in the work of Ref. \cite{haya1}, a negative mass-shift of magnitude $50$ MeV
was observed. The properties of pseudoscalar $B$ mesons as well
as charm strange $D_s$ mesons were also studied in Ref. \cite{higler1}
and these studies were further extended to the scalar $D$ mesons 
in Ref. \cite{higler2}.
 In Ref. \cite{higler3}, the 
OPE for pseudoscalars $D$ and $B$ mesons was
performed
for both, the normal-order operators and also for non-normal-ordered operators.
A projection method to calculate 
the higher order contributions to OPE, for 
in-medium QCD sum rules of $D$ mesons was used
in Ref. \cite{higler4}. A study on the decomposition
of higher dimensional condensates into vacuum part and a medium part was
done recently in \cite{higler5}.
The heavy quark expansion for the study of $D$ and $B$ mesons 
in the nuclear matter was performed in \cite{higler6}.
The Wilson coefficients and four quark condensates
in the QCD sum rules for the medium modification of
$D$ mesons were calculated in Ref. \cite{higler7}.

The properties of open and hidden charm mesons have also been studied to great
extent in coupled channel approach \cite{tolosint1,tolosint3,tolosint4,tolosint5,tolosint6,tolosint7,
tolosint8,tolosint9,
tolosint2}.
Using the separable potential technique between
meson-baryon interactions and SU(3) symmetry
among the $u$, $d$ and $c$ quarks, the  spectral
properties of $D$ mesons were investigated in the nuclear
matter at zero \cite{tolosint3} and finite temperatures \cite{tolosint4}.
In the work of references \cite{tolosint5,tolosint6}, the properties of 
$D$ mesons were studied in SU(4) sector including the strangeness channel
also and breaking SU(4) symmetry through the exchange of vector
mesons.  In Ref. \cite{tolosint7}, the properties
of $D$  and $\bar{D}$ mesons were investigated
self-consistently in coupled channel approach including
the Pauli blocking effects and mean field potential of baryons.
The properties of scalar charm resonances and hidden charm resonances
were also studied using coupled channel method in \cite{tolosint8}. 
The study of open charm mesons
within coupled channel approach was further
improved by using heavy quark spin-flavour symmetry
and the in-medium properties of charmed pseudoscalar
and vector  mesons
as well as charm baryon resonances $\Lambda_c$, $\Sigma_c$, $\Xi_c$ and
$\Omega_c$  were investigated \cite{tolosint91,tolosint9} .
In Ref. \cite{tolosint1}, the chiral unitary coupled channel
approach was used to study the interactions of
light vector mesons with baryon and nuclei
and the possibilities of the formation of quasibound states of vector mesons with
baryons were observed. These studies were also extended to charm sector
where hidden charm states were observed and which
has consequences on $J/\psi$ suppression \cite{tolosint1}.
The baryons states with charm degree of freedom can be generated
dynamically within the coupled channel approach and
also incorporating the heavy quark spin symmetry \cite{tolosint2}.

 The chiral hadronic SU(3) model
 is generalized to SU(4) and SU(5) sector,
 for investigating the in-medium properties of pseudoscalar
 $D$ and $B$ mesons \cite{amavstranged,amdp1,amdp2}.
The in-medium mass modifications of scalar, vector 
and axial-vector heavy
$D$ and $B$ mesons were investigated using
 the QCD sum rules in the nuclear matter in \cite{wang1,wang2}.
 In Ref. \cite{arv1}, we studied the mass-modifications of 
 scalar, vector and axial vector heavy charmed and bottom mesons
 at finite density of the nuclear matter using the chiral SU(3) model
 and QCD sum rules. In this approach,
 we evaluated the medium modifications of quark and gluon condensates 
 in the nuclear matter through the 
 medium modification of scalar isoscalar fields
 $\sigma$ and $\zeta$ and the scalar dilaton field $\chi$.
In the present paper, we shall evaluate the properties
of vector and axial-vector mesons
 in the hot and dense isospin asymmetric strange hadronic medium. 
In the QCD sum rules, the properties of above mesons
are modified through the quark and gluon condensates.
Within the chiral hadronic model the
 quark and gluon condensates are written 
 in terms of scalar fields
$\sigma$, $\zeta$ and $\delta$ and the
scalar dilaton field $\chi$. We shall evaluate the $\sigma$, 
$\zeta$, $\delta$ and $\chi$ fields
and hence the quark and gluon condensates 
in the medium consisting of nucleons and hyperons
 and shall find the 
mass shift and shift in decay constants of heavy vector
 and axial vector mesons.
 
The study of decay constant of heavy mesons
 play important role in understanding the
 strong decay of heavy mesons, their electromagnetic
 structure as well radiative decay width.
The study of B meson decay constants
is important for $B_d$-$\bar{B}_d$
and $B_s$-$\bar{B}_s$ 
mixing \cite{ebert1}.
The  decay constants of vector $D^{*}$
and $B^{*}$ mesons 
are helpful for calculations of strong coupling
in $D^{*}D\pi$ and $B^{*}B\pi$
mesons using light cone sum rules \cite{bel1}.
An extensive literature is available on
the calculations of decay constants
of heavy mesons in the free space,
for example, the QCD sum
rules based on OPE of
two-point correlation function,
heavy-quark expansion \cite{shurya1}, sum rules
in heavy-quark effective theory \cite{heq1,heq2} and
sum rules with gluon radiative corrections
to the correlation functions upto
two loop \cite{gel2,gel3,gel4} or three loop \cite{gel1}.
However, the in-medium modifications
of  decay constants
of heavy mesons had been studied
 in symmetric nuclear matter
only \cite{azzi,wang3}.
  The thermal modification (at zero baryonic density) of 
 decay constants of heavy vector mesons 
 was investigated in Ref. \cite{azzi2}
 and it was observed  that the values of decay constants remained almost
 constant upto 100 MeV, but above this
 decrease sharply with 
 increase in temperature.

We shall present this work as follows:
In Sec. II, we shall briefly describe the chiral SU(3) model 
which is used to evaluate the quark and gluon condensates 
in the strange hadronic matter. Section III will introduce 
the QCD sum rules which we shall use in the present 
work along with chiral model to evaluate the
in-medium properties of mesons. In Sec. IV, we shall present
our results of present investigation and possible discussion
on these results. Section V will summarize the present work.

 \section{Chiral SU(3) model}
The basic theory of strong interaction, the QCD, is not 
directly applicable in the non-perturbative regime.
To overcome this limitation, the effective theories,
constrained by the basic 
properties, e.g., chiral symmetry and scale invariance of
QCD, are constructed. The chiral SU(3) model is one such effective model
 based on the non-linear realization
 and broken scale invariance as well as
 spontaneous breaking  properties of chiral symmetry
  \cite{paper3,amarvindjpsi}.
 The glueball field $\chi$ is introduced in the model
  to account for the broken scale invariance properties of QCD.
  The model
 had been used successfully in the literature  to study the
 properties of hadrons at finite density and temperature of the
 nuclear and strange hadronic  medium. 
 The 
 general Lagrangian density of the chiral SU(3) model 
 involve the kinetic energy terms, the baryon meson interactions, self
 interaction of vector mesons, scalar mesons-meson 
 interactions as well as the explicit
 chiral symmetry breaking term \cite{paper3}.
 From the Lagrangian densities
  of the chiral SU(3) model, using
the mean field approximation, we find the coupled equations of motion
for the scalar fields $\sigma$, $\zeta$, $\delta$ and  the
scalar dilaton field $\chi$ \cite{amarvindjpsi}.
These coupled equations of motion for the above fields are solved 
for the
different values of isospin asymmetry parameter $I$, and strangeness
fraction $f_s$, of the hadronic medium at
finite density and temperature. The isospin asymmetry
parameter, $I$, is defined by the relation, 
$I = \frac{\rho_n - \rho_p}{2\rho_B}$, 
where, $\rho_n$ and $\rho_p$, are
the number densities of neutrons and protons respectively
and $\rho_B$ is baryon density.
The strangeness fraction of the medium is 
given by, $f_s =  \frac{\sum_i |s_i| \rho_i}{\rho_B}$.
Here, $s_i$, is the number of strange quarks
and $\rho_i$ is the number density of $i^{th}$ baryon \cite{amarvindjpsi}.

In the present work, for the
 evaluation of vector and axial-vector
meson properties using QCD sum rules, we shall need the
light quark condensates $\langle\bar{u}u\rangle$ and $\langle\bar{d}d\rangle$, 
the strange quark condensate $\langle\bar{s}s\rangle$ and the
scalar gluon condensate $\langle \frac{\alpha_s}{\pi}G_{\mu\nu}^{a} G^{\mu\nu a}\rangle$. 
In the chiral effective model, the explicit symmetry breaking term
 is introduced to
eliminate the Goldstone bosons
  and can be
used to extract the scalar quark condensates, $\langle q\bar{q}\rangle$, 
in terms of scalar fields $\sigma$, $\zeta$, $\delta$ and $\chi$.
We write \cite{amarvindjpsi},
 \begin{eqnarray}
&&\sum_i m_i \bar {q_i} q_i = - {\cal L} _{SB} \\
& =& \left( \frac {\chi}{\chi_{0}}\right)^{2} 
\left( \frac{1}{2} m_{\pi}^{2} 
f_{\pi} \left( \sigma + \delta \right) +
\frac{1}{2} m_{\pi}^{2} 
f_{\pi} \left( \sigma - \delta \right)
 + \big( \sqrt {2} m_{k}^{2}f_{k} - \frac {1}{\sqrt {2}} 
m_{\pi}^{2} f_{\pi} \big) \zeta \right). 
\label{quark_expl}
\end{eqnarray}
 From Eq. (\ref{quark_expl}), light  
 scalar quark condensates $\left\langle\bar{u}u\right\rangle $ and
  $\left\langle\bar{d}d\right\rangle$, and the strange quark
 condensate $\left\langle  \bar{s}s\right\rangle $,
  can be written as,
 \begin{equation}
\left\langle \bar{u}u\right\rangle 
= \frac{1}{m_{u}}\left( \frac {\chi}{\chi_{0}}\right)^{2} 
\left[ \frac{1}{2} m_{\pi}^{2} 
f_{\pi} \left( \sigma + \delta \right) \right],
\label{qu}
\end{equation}
\begin{equation}
\left\langle \bar{d}d\right\rangle 
= \frac{1}{m_{d}}\left( \frac {\chi}{\chi_{0}}\right)^{2} 
\left[ \frac{1}{2} m_{\pi}^{2} 
f_{\pi} \left( \sigma - \delta \right) \right],
\label{qd}
\end{equation}
and
\begin{equation}
\left\langle \bar{s}s\right\rangle 
= \frac{1}{m_{s}}\left( \frac {\chi}{\chi_{0}}\right)^{2} 
\left[ \big( \sqrt {2} m_{k}^{2}f_{k} - \frac {1}{\sqrt {2}} 
m_{\pi}^{2} f_{\pi} \big) \zeta \right],
\label{qs}
\end{equation}
respectively.
Also, we know that, due to broken scale invariance property of QCD, 
 the trace of energy momentum tensor is
non-vanishing (trace anomaly) and is equal to the scalar gluon condensates. The trace anomaly property of QCD can 
be mimicked in effective chiral model
through the scale  breaking Lagrangian density
which can be further used to evaluate the 
trace of energy momentum tensor. Comparing, the trace of energy momentum tensor evaluated from effective chiral
model, to the trace of energy  momentum tensor of
basic QCD theory, we can express the gluon condensate
in terms of scalar fields $\sigma$, $\zeta$, $\delta$ and
$\chi$ \cite{paper3,amarvindjpsi}.
This relation is given by, 
\begin{equation}
\left\langle  \frac{\alpha_{s}}{\pi} {G^a}_{\mu\nu} {G^a}^{\mu\nu} 
\right\rangle =  \frac{8}{9} \Bigg [(1 - d) \chi^{4}
+\left( \frac {\chi}{\chi_{0}}\right)^{2} 
\left( m_{\pi}^{2} f_{\pi} \sigma
+ \big( \sqrt {2} m_{k}^{2}f_{k} - \frac {1}{\sqrt {2}} 
m_{\pi}^{2} f_{\pi} \big) \zeta \right) \Bigg ]. 
\label{chiglu}
\end{equation}
\Cref{qu,qd,qs,chiglu} give us the expressions for different
scalar quark condensates and gluon condensate in terms
of scalar fields within chiral SU(3) model.
It should be noted that, for massless quarks, 
 the second term in Eq. (\ref{chiglu}), 
arising from explicit symmetry breaking will be absent and  the scalar 
gluon condensate becomes proportional to the fourth power of the 
dilaton field $\chi$, in the chiral SU(3) model.

 \section{QCD sum rules for vector and axial-vector heavy mesons in strange
 hadronic matter}
 In this section, we shall discuss the QCD sum rules \cite{wang1,wang2}, to be
used later along with the chiral SU(3) model, for the evaluation of 
in-medium properties of vector and axial-vector mesons
in asymmetric strange hadronic matter. 
 We know that, the starting point in the QCD sum rules is to write 
 the two-point correlation function, $\Pi_{\mu\nu}(q)$, which is the
 Fourier transform 
of the expectation value of 
time ordered product of isospin averaged currents $J_\mu(x)$  and 
can be decomposed into  vacuum 
part, a static nucleon part and pion bath contribution \cite{arv1,kwon1,zscho1}.  
 For the vector and axial-vector mesons the average particle-antiparticle currents are given by, 
$ J_\mu(x) = J_\mu^\dag(x) =\frac{\bar{c}(x)\gamma_\mu q(x)+\bar{q}(x)\gamma_\mu c(x)}{2}$
 and
 $ J_{5\mu}(x) = J_{5\mu}^\dag(x) =\frac{\bar{c}(x)\gamma_\mu \gamma_5q(x)+\bar{q}(x)\gamma_\mu\gamma_5 c(x)}{2}$,
respectively.
Here, $q$, denotes the  $u$, $d$ or $s$ quark
 (depending upon type of  meson under investigation), whereas, $c$, denotes the heavy charm quark
 (for $B$ mesons the quark, $c$, will be replaced by bottom, $b$, quark).
We know that the charmed mesons $D^{+}$, $D^{-}$, $D^{0}$
and $\bar{D^{0}}$ have the quark compositions,
$c\bar{d}$, $d\bar{c}$, $c\bar{u}$ and $u\bar{c}$,
respectively. The mesons
$D^{+}$ and  $D^{-}$ are particle-antiparticles to each-other
  and similarly, $D^{0}$
and $\bar{D^{0}}$ mesons. The above listed charm mesons belong to
the isospin, $D (D^{+}, D^{0})$ and $\bar{D} (D^{-}, \bar{D^{0}})$ doublets.
 In the present work, we shall find the average mass-shift for
 particle-antiparticle pairs, i.e., for, $D^{+}$ and $D^{-}$,
and for,  $D^{0}$ and  $\bar{D^{0}}$, under centroid approximation \cite{haya1,wang2}.
In this way, we will also be able to study the 
 mass-splitting of isospin doublet 
 due to isospin asymmetry of the medium.
To find the mass splitting of particles
and antiparticles in the nuclear medium one has to consider the even and 
odd part of QCD sum rules \cite{higler1}.
For example, in Ref. \cite{higler1} the mass splitting between 
pseudoscalar $D$
and $\bar{D}$ mesons was investigated using the even and odd QCD sum rules,
whereas in \cite{haya1,wang1,wang2} the mass-shift of $D$ mesons was investigated
under centroid approximation. It means, in the present work we
assumed the degeneracy between particle-antiparticle mass. 
Note that the sum rules for $D^{0}$
can be obtained from the $D^{+}$ just by replacing $d$ quark by $u$ quark \cite{higler3}. 

  In literature of QCD sum rules,  
  the pion bath contributions are
  used to evaluate the effects of finite temperature of the
  medium on the properties of mesons \cite{kwon1,zscho1}. In our present approach,
  however, the effects of
  finite temperature of the medium on the properties of
  vector and axial-vector mesons are
  evaluated through the temperature dependence of
  scalar fields $\sigma$, $\zeta$, $\delta$ and $\chi$, 
  which will appear through static one nucleon part \cite{isoamss,amarvind,amavstranged,amarvindjpsi,arv1}.
For the case of vector and axial vector mesons, the correlation function $T^{N}_{\mu\nu}(\omega,\mbox{\boldmath $q$}\,)$, which appear in the static nucleon part,
can be written in terms of forward scattering amplitudes
$T_N(\omega,\mbox{\boldmath $q$}\,)$ (corresponding to vector
and axial-vector mesons) and 
$T_{N}^{0} (\omega, \textbf{q})$ (corresponding to
scalar and pseudoscalar mesons), as follows
 \cite{koike1,wang3}, 
\begin{eqnarray}
T^{N}_{\mu\nu}(\omega,\mbox{\boldmath $q$}\,) = T_{N} (\omega, \textbf{q})
\left(\frac{q_{\mu} q_{\nu}}{q^2 } - g_{\mu \nu} \right) +
T_{N}^{0} (\omega, \textbf{q})\frac{q_{\mu} q_{\nu}}{q^2 } .
\label{spectraln1}
\end{eqnarray}
Note that in the present work, we are interested in the 
properties of vector and axial-vector mesons and therefore, second term of Eq. (\ref{spectraln1}) will not contribute. The forward scattering amplitude $T_{N} (\omega, \textbf{q})$,
in the limit, $\textbf{q}\rightarrow 0$, can be written in
terms of spin averaged spectral density $\rho$, as follows \cite{koike1, azzi},
\begin{eqnarray}
T_{N} (\omega, \textbf{q}) = \int_{-\infty}^{+\infty} du \frac{\rho(u, \textbf{q} = 0)}{u - \omega -i \epsilon} = \int_{0}^{\infty} du^2 \frac{\rho(u, \textbf{q} = 0)}{u^2 - \omega^{2}}.
\label{spectraln2}
\end{eqnarray}
Near the pole positions of vector and axial vector mesons,
the phenomenological spectral densities can be
 further related to the $D^{*}N$ and $D_{1}N$
scattering  matrix ${\cal T}$,
and can be
 parametrized with three unknown parameters
$a, b$ and $c$ as discussed in Refs. \cite{wang1,wang2,haya1,arv1}.
Also, the in-medium mass shift, $\delta m_{D^*/D_1}$, of vector and axial-vector mesons can be related to their scattering
length $a_{D^*/D_1}$, through parameter $a$ and is given by  \cite{wang1,wang2,haya1,arv1},
\begin{eqnarray}
\delta m_{D^*/D_1} &=& \sqrt{m_{D^*/D_1}^2+
\left(\frac{\rho_B }{2 m_N} \frac{a }{f_{D^*/D_1}^2m_{D^*/D_1}^2}\right) }
- m_{D^*/D_1} \nonumber\\
&=& \sqrt{m_{D^*/D_1}^2+
\left(-\frac{\rho_B }{2m_N}8\pi(m_N+m_{D^*/D_1})a_{D^*/D_1}\right) }
- m_{D^*/D_1} .
\label{massshift}
\end{eqnarray} 
The shift in the decay constant of vector or axial vector mesons 
can be written as \cite{wang3},
\begin{eqnarray}
\delta f_{D^{*}/ D_1} =  \frac{1}{2f_{D^{*}/D_1}m_{D^{*}/D_1}^{2}}
\left(\frac{\rho_{B}}{2m_{N}}b - 
2f_{D^{*}/D_1}^{2} m_{D^{*}/D_1}
\delta m_{D^{*}/D_1} \right) .
\label{decayshift}
\end{eqnarray}

From  Eqs.
 (\ref{massshift}) and (\ref{decayshift}) we observe 
 that,  to find the value of mass shift and shift in decay constant
 of mesons, we first need to know the value of unknown parameters $a$ and $b$.
For achieving this, the equations for  Borel transformation of the scattering matrix on the
phenomenological side are equated  to the Borel transformation of the scattering matrix for the 
OPE side \cite{haya1,wang1,wang2}.
The detailed expressions for the Borel transformed
equations of heavy vector and axial-vector mesons
can be found in Refs. \cite{wang2,arv1}.
The unknown parameters $a$ and $b$, will appear in these
Borel transformed equations.
To obtain the values of these
two unknown parameters, the Borel transformed equation of the meson under study is
differentiated  w.r.t.
$\frac{1}{M^{2}}$,  so that we could have two equations and two unknowns. By solving those two coupled equations we will be able to get the
values of parameters, $a$ and $b$. 

Also note that, the Borel transformed equations of vector and axial-vector mesons will
depend on the nucleon expectation values of quark
condensates $ \left\langle \bar{q}q\right\rangle$, 
$\langle q^\dag i D_0q\rangle$,
$\langle\bar{q}g_s\sigma Gq\rangle$
and $\langle \bar{q} i D_0 i D_0q\rangle$,
 and the gluon condensate $\left\langle  \frac{\alpha_{s}}{\pi} {G^a}_{\mu\nu} {G^a}^{\mu\nu} 
\right\rangle$ \cite{arv1}.
The nucleon expectation values of the operators can be expressed in terms of
the expectation value of operators at finite baryonic density and vacuum expectation value \cite{koike1,zscho1,kwon1,arv1}.
As discussed earlier, in Sec. II, the scalar quark condensate $ \left\langle \bar{q}q\right\rangle$ and the gluon condensates
$\left\langle  \frac{\alpha_{s}}{\pi} {G^a}_{\mu\nu} {G^a}^{\mu\nu} 
\right\rangle$, 
can be evaluated within chiral SU(3) model.
The condensates, $\langle\bar{q}g_s\sigma Gq\rangle$
and $\langle \bar{q} i D_0 i D_0q\rangle$,
can be approximated in terms of scalar quark condensates
$ \left\langle \bar{q}q\right\rangle _{\rho_{B}} $ \cite{qcdThomas}.
 The value of $ \left\langle \bar{q}q\right\rangle$,
 evaluated within chiral SU(3) model,
can be used in the equations of
$\langle\bar{q}g_s\sigma Gq\rangle$
and $\langle \bar{q} i D_0 i D_0q\rangle$,
for an improved evaluation of these condensates within chiral model. 
 The value of quark condensate, $\langle q^\dag i D_0q\rangle$, for light quark is equal to  $0.18 GeV^2 \rho_{B}$ and for strange quark it is $0.018 GeV^2 \rho_B$ \cite{qcdThomas}.
 
Note that, the bottom mesons, i.e.,
$B^{+}$, $B^{-}$, $B^{0}$
and $\bar{B^{0}}$, have the quark compositions,
$u\bar{b}$, $b\bar{u}$, $d\bar{b}$ and $b\bar{u}$, respectively.
Comparing the quark compositions of charm and bottom mesons,
e.g., for $B^{+}$ and $\bar{D^{0}}$, we observe that
both have same light $u$ quark and differ in 
 heavy flavor antiquark. 
In the QSR equations of
 $D$ mesons, 
 the quark condensates corresponding to only light quark appear
 along with the masses of heavy quark
 and mesons. It means QSR equation of
 $B^{+}$ will be differ from that of $\bar{D^{0}}$
 mere by masses of heavy flavor mesons (both have same light quark).
Same correspondence exist for
 $B^{-}$ and $D^{0}$, $B^{0}$ and $D^{-}$
 and $\bar{B^{0}}$ and $D^{+}$. Thus sum rules for
 the bottom $B$ mesons can be obtained from the sum rules of
 $D$ mesons by replacing,
 $m_c$ by $m_b$,
$D^{*}$ ($D_1$) by $B^{*}$ ($B_1$)
and also by replacing $\Lambda_c$ ($\Sigma_c$) by 
$\Lambda_b$ ($\Sigma_b$).
The same strategy was applied
to obtain the OPE for pseudoscalar $B$
mesons from $D$ mesons in Refs. 
\cite{higler3,wang1,wang2,wang3,azzi}. 
When we move from non-strange to
strange mesons, the light quark is replaced by strange quark. 
Therefore, the QCD sum rules for the strange  vector mesons are obtained by
using the strange condensates in place of light quark condensates.
Hilger in Ref. \cite{higler1} used the same strategy for 
obtaining the sum rules for pseudoscalar 
$D_s$ mesons  from $D$ meson sum rules.
Now the sum rule equations of $D_s$ mesons will also be
differ from $B_s$ mesons by the mass of heavy flavor
mesons only and therefore, sum rules for strange $B_s$ mesons are
obtained by replacing masses of $D_s$ mesons by corresponding
 masses of $B_s$
mesons.

 \section{Results and Discussions}
In this section, we shall discuss the results of present
investigation of in-medium mass shift and shift in decay 
constant of 
vector  [$D^{*} (D^{* +}, D^{* 0}, D_{s}^{*})$ and  
$B^{*} (B^{* +}, B^{* 0}, B_{s}^{*})$] and
axial-vector [$D_{1} (D^{+}_{1}, D^{0}_{1}, D_{1s})$ and 
$B_{1} (B^{+}_{1}, B^{0}_{1}, B_{1s})$] mesons in
isospin asymmetric strange hadronic matter. 
First we list the values of various parameters
used in the present work on vector and
axial-vector mesons. 
Nuclear matter saturation density adopted in 
the present investigation is 0.15 $fm^{-3}$. 
The coupling constants of heavy baryons with
 nucleons and heavy charmed/bottom mesons 
appear in the hadronic matrix elements of
the heavy baryon transitions to nucleon. 
These coupling constants are calculated
using the light cone QCD sum rules \cite{khod}.
The average value of coupling constants,
$g_{{D^*N\Lambda_c}}$ and $g_{{D^*N\Sigma_c}}$ is $3.86$ \cite{khod,wang2,wang3}.
As we shall discuss later, the shift in masses and decay constants of
vector and 
axial-vector mesons are not much sensitive to the
values of coupling constants and therefore, we
used the same values for all vector
and axial vector mesons under investigation \cite{wang2,wang3}, i.e.,
we use,
$g_{{D^*N\Lambda_c}}$ $\approx$ $g_{{D^*N\Sigma_c}}$ $\approx$ $g_{{D_1N\Lambda_c}}$ $\approx$ $g_{{D_1N\Sigma_c}}$ $\approx$ $g_{{B^*N\Lambda_b}}$ $\approx$ $g_{{B^*N\Sigma_b}}$ $\approx$  $g_{{B_1N\Lambda_b}}$ $\approx$ $g_{{B_1N\Sigma_b}}$
 equal to 3.86 \cite{wang2}. 
  The masses of quarks namely, up ($u$), down ($d$), strange ($s$),
   charm  ($c$), and bottom ($b$), used
  in the present work 
  are, 0.005, 0.007, 0.095, 1.35 and 4.7 GeV, respectively.
 
 In Table \ref{table_massffss},
  we mentioned the values of masses $m$, decay constants $f$ and threshold parameters $s_0$,
 of vector and axial-vector mesons \cite{wang2,wang3}.
      \begin{table}
\begin{tabular}{ |c|c|c|c|c|c|c|c|c| }
\hline
  & ${D^{*+}}$ & ${D^{*0}}$ &  ${B^{*0}}$ & ${B^{*+}}$ & ${D_s^{*}}$ & ${B_s^{*}} $ \\
 \hline
 $ m $ (GeV) & 2.01 & 2.006 & 5.325 & 5.323 & 2.112 & 5.415 \\  
 \hline       
 $ f $ (GeV) & 0.270 & 0.270 & 0.195 & 0.195 & 1.16$f_{D^*}$  & 1.16$f_{B^*}$ \\  
 \hline       
 $ s_0 $ (GeV$^2$)  & 6.5 & 6.5 &  35 & 35 & 7.5 & 38 \\  
 \hline \hline  
  & ${D_1^{+}}$ & ${D_1^{0}}$ & ${B_1^{0}}$ & ${B_1^{+}}$ &  ${D_{1s}} $ & ${B_{1s}} $ \\
 \hline
 $ m $ (GeV)  & 2.423 & 2.421 & 5.723 & 5.721 & 2.459 & 5.828 \\  
 \hline   
 $ f $ (GeV)  & 0.305 & 0.305 &  0.255 & 0.255 &1.16$f_{D_1}$ & 1.16$f_{B_1}$ \\  
 \hline   
  $ s_0$  (GeV$^2$) & 8.5 & 8.5 & 39 & 39 & 9.5 & 41 \\  
 \hline 
 \end{tabular}
\caption{In the above table, we tabulate the values of  masses $m$, decay constants $f$ and threshold parameters $s_0$, of vector and axial-vector mesons.}
\label{table_massffss}
\end{table}
The values  of decay constants of
strange vector mesons are calculated in
terms of decay constants of non-strange vector
mesons in Refs. \cite{damir,bakers,gel1,hwang}.
In Ref. \cite{gel1}, using QCD sum rules, 
the values for $f_{{D_s}^*}$ and $f_{{B_s}^*}$
 are found to be $1.21 f_{{D}^*}$ and $1.20 f_{{B}^*}$, respectively.
 In \cite{damir}, the decay constants of $f_{{D_s}^*}$ are
calculated from
the simulations of twisted mass QCD and is observed to be $1.16 f_{{D}^*}$.
The values of $f_{{B_s}^*}$ in \cite{hwang}  is
observed to be $1.16 f_{{B}^*}$. 
 In our present calculations, for
 strange vector and axial-vector mesons, we used the values $1.16$ times
 the decay constant of respective non-strange meson,
i.e., the values of $f_{D_s^*}$, $f_{B_s^*}$, $f_{D_{1s}}$ and $f_{B_{1s}}$ are $1.16 f_{{D}^*}$, $1.16 f_{{B}^*}$, $1.16 f_{D_{1}}$, 
and $1.16 f_{B_{1}}$, respectively \cite{damir,hwang}.
The continuum threshold parameter $s_0$, define the scale below which
the continuum contribution
vanishes \cite{kwon1}.
       The values of continuum threshold parameters $s_0$ are chosen so as
to reproduce  the experimental observed vacuum values of
the masses of vector and axial-vector mesons \cite{wang2,wang3}. 
      We assume that the various coupling constants and 
          continuum threshold parameters are not subjected to 
          medium modifications. 
          
To describe the exact mass (decay) 
          shift of above mesons we have chosen a suitable Borel
           window, i.e., the range of squared Borel mass
           parameter $M^{2}$, 
           within which there is almost no variation in the mass and
            decay constant.  
            In Table \ref{table_parameterss}, 
            we mentioned the Borel windows,
            as observed in the present calculations for the
            mass shift and shift in decay constant of vector and axial-vector 
            mesons. 
            In our calculations,
we observe that, masses
and decay constant of vector and axial-vector mesons
are stable in different regions of $M^2$ and 
therefore, different Borel windows for masses and decay constant
of respective mesons are mentioned in Table \ref{table_parameterss}.                                                                                                                                                                                                                                                                                                                                                                                                                                                                                                                                                                                                                                                                                                                                                                                                                                                                                                                                                                                                                                                                                                                                                                                                                                                                                                                                                                                                                                                                                                                                                                                                                                                                                                                                                                                                                                                                                                                                                                                                                                                                                                                                                                                                                                                                                                                                                                                                                                                                                                                                                                                                                                                                                                                                                                                                                                                                                                                                                                                                                                                                                                                                                                                                                                                                                           This is consistent with observations of 
Ref. \cite{wang3}. Our further discussion in the present section is divided into three subsections.
In subsection $A$, we shall discuss about the
behavior of scalar fields and condensates in the strange hadronic medium, at zero and finite temperature. Subsections $B$ and $C$ will be devoted to present the results on vector and axial-vector mesons, respectively.
\begin{table}
\begin{tabular}{ |c|c|c|c|c|c|c|c|c| }
\hline
  & ${D^{*+}}$ & ${D^{*0}}$ &  ${D_s^{*}} $ & ${B^{*+}}$ & ${B^{*0}}$ & ${B_s^{*}} $ \\
 \hline
 $\delta m$ & (4.5 - 6.5) & (4.5 - 6.5) & (5.0 - 7.0) & (30 - 33) & (30 - 33) & (31 - 34) \\  
 \hline       
 $\delta f$  & (3.3 - 4.9) & (3.3 - 4.9) & (3.8 - 5.3) & (26 - 31) & (26 - 31) & (27 - 31) \\  
 \hline \hline 
  & ${D_1^{+}}$ & ${D_1^{0}}$ &  ${D_{1s}} $ & ${B_1^{+}}$ & ${B_1^{0}}$ & ${B_{1s}} $ \\
 \hline
 $\delta m$  & (5.4 - 9.4) & (5.4 - 9.4) & (5.9 - 9.9) & (33 - 38) & (33 - 38) & (36 - 40) \\  
 \hline   
 $\delta f$  & (4.2 - 7.2) & (4.2 - 7.2) & (5.0 - 8.0) & (30 - 34) & (30 - 34) & (32 - 36) \\  
 \hline
   \end{tabular}
\caption{In the above table, we tabulate the Borel windows i.e. range of squared Borel mass parameter $M^2$,  within which shift in mass $\delta m$ and decay constant  $\delta f$ remain stable. The values shown in table are in GeV$^2$ units and are given for both vector and axial-vector mesons.}
\label{table_parameterss}
         \end{table} 
           
       \subsection{Scalar Fields and Condensates:}      
We start with the discussion on the behavior of
quark and gluon condensates for different strangeness fractions
 and isospin asymmetry parameters
 of strange hadronic medium. 
In literature, the quark condensates are evaluated to
leading order in nuclear density
using the Feynman Hellmann theorem
and model independent results were obtained 
in terms of pion nucleon sigma term \cite{cohen, quark2}.
Using the Feynman Hellmann theorem, the quark
condensate at finite density of nuclear matter
is expressed as a sum of vacuum value and a term dependent
on energy density of nuclear matter.
In the model independent calculations,
the interactions between nucleons were 
neglected and free space nucleon mass was used.
If one use only the leading order calculations
for the evaluation of
quark condensates above
nuclear matter density,
then the quark condensates decreases very sharply 
and almost vanishes around 3$\rho_0$.
In Ref. \cite{quark3}, the Dirac-Brueckner approach
with the Bonn boson-exchange potential
was used to include the higher order corrections
and to find the quark condensates in the
nuclear matter above the
nuclear matter density. The calculations show that
at higher density the quark condensates
decrease more slowly as compared to
leading order predictions.

\begin{figure}
\includegraphics[width=16cm,height=14cm]{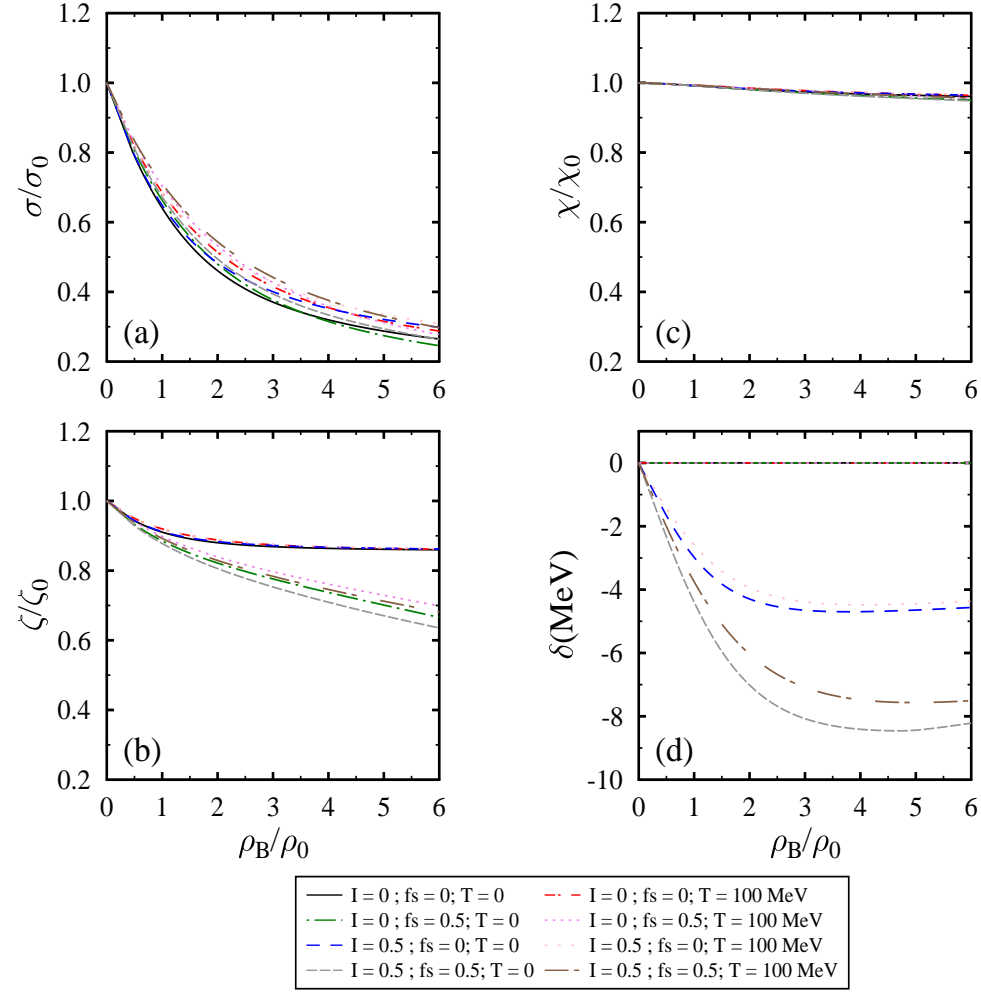}  
\caption{(Color online)
In the above figure, subplots (a), (b) and (c) show 
the variation of ratio of in-medium value  to vacuum value
of scalar fields, $\sigma/\sigma_0$, $\zeta/\zeta_0$
and $\chi/\chi_0$
 as a function of
 baryonic density, $ \rho_B$ (in units of nuclear saturation density, $\rho_0$). 
 In subplot
 (d) we have shown the scalar-isovector field $\delta$
 as a function of density of medium.  We show the results at temperatures,
 $T$ = 0 and 100 MeV. 
 For each value of temperature, the
 results are plotted for isospin asymmetry parameters, $I$ = 0
 and 0.5 and the
 strangeness fractions, $f_s$ =0 and 0.5.}
\label{fields}
\end{figure}

As discussed earlier, in the chiral SU(3) model, used in the present work, the
quark and gluon condensates are expressed in terms of 
scalar fields $\sigma$, $\zeta$, $\delta$ and
 $\chi$ \cite{amarvind,amarvindjpsi}.
In 
Fig. \ref{fields}, 
we show the
variation of, ratio of in-medium value to vacuum value
of scalar fields, as a function of baryonic
density of strange hadronic medium. We show
the results for  temperatures $T = 0$  and
$T = 100$ MeV.
For each value of temperature $T$,
the results are plotted for
isospin asymmetry parameters
$I$ = 0  and 0.5 and the strangeness
fractions $f_s$  = 0 and 0.5.
From Fig. \ref{fields}, we can see that, the  scalar fields
$\sigma$ and $\zeta$ varies considerably
as a function of baryonic density of
medium, whereas the scalar field $\chi$
has little density dependence. 
For example, in symmetric nuclear medium ($I$ = 0 and $f_s$ = 0),
at density $\rho_B$ = $\rho_0$ ($4\rho_0$),
the values of scalar fields $\sigma$, $\zeta$ and $\chi$,
are observed to be
0.64 $\sigma_0$ (0.31 $\sigma_0$), 
0.91 $\zeta_0$ (0.86 $\zeta_0$), and
0.99 $\chi_0$ (0.97 $\chi_0$). The symbols,
$\sigma_0$, $\zeta_0$ and $\chi_0$ denote the vacuum values of
scalar fields and have values
$-93.29$, $-106.75$ and $409.76$ MeV, respectively.
It is observed
that, at high baryon density, the strange scalar-isoscalar field
$\zeta$, varies considerably as a function of
strangeness fraction $f_s$, as compared to 
non-strange scalar isoscalar field $\sigma$.
For example, in symmetric medium ($I$ = 0), at baryon density
$4\rho_0$,
as we move from $f_s$ = 0 to $f_s$ = 0.5,
the value of $\zeta$ changes by  14 \%, whereas the
value of $\sigma$ changes by 1\% only.
However, the effect of isospin asymmetry
of the medium is more 
on the values of $\sigma$ field as
compared to $\zeta$ field. For example, in nuclear medium ($f_s$ = 0), at
baryon density $\rho_B$  = 4$\rho_0$,
as we move from $I$  = 0 to 0.5, the values of
non-strange scalar field $\sigma$ and the strange scalar
field  $\zeta$ changes by 10.25\% and 0.24 \% respectively.
However, in strange medium, at $f_s$ = 0.5,
the percentage change in the values of $\sigma$
and $\zeta$ is 7.2\% and 7\% respectively.
Since the scalar meson $ \sigma$,
has light quark content ($u$ and $d$ quarks)  and the $\zeta$ meson
have strange quark content ($s$ quark) and therefore, former
is more sensitive to 
isospin asymmetry of the  medium 
(property of $u$ and $d$ quarks) and the
latter is to the
strangeness fraction.
For the finite baryonic density of the
medium,  the scalar fields undergo less drop 
at non-zero temperature of the medium, as compared
to zero temperature situation.
 This effect 
of temperature is more pronounced at 
finite strangeness and higher baryonic density
of the medium. For example, in symmetric 
nuclear matter, at baryonic density
 $\rho_B$ = $\rho_0$,  for a change of temperature
from $T = 0$ to $T = 100$ MeV, the magnitude of scalar field  $\sigma$,
is increased by approximately $6\%$. 
However, at $\rho_B$ = $4\rho_0$ and $f_s = 0.5$, the
increase in the magnitude of $\sigma$ is $12\%$.
The observed behavior of scalar fields with the temperature
of the medium, in the present chiral SU(3) 
hadronic model, is consistent with
the calculation within chiral quark mean field model
discussed in Ref. \cite{quarkmean1}. 

\begin{figure}
\includegraphics[width=16cm,height=16cm]{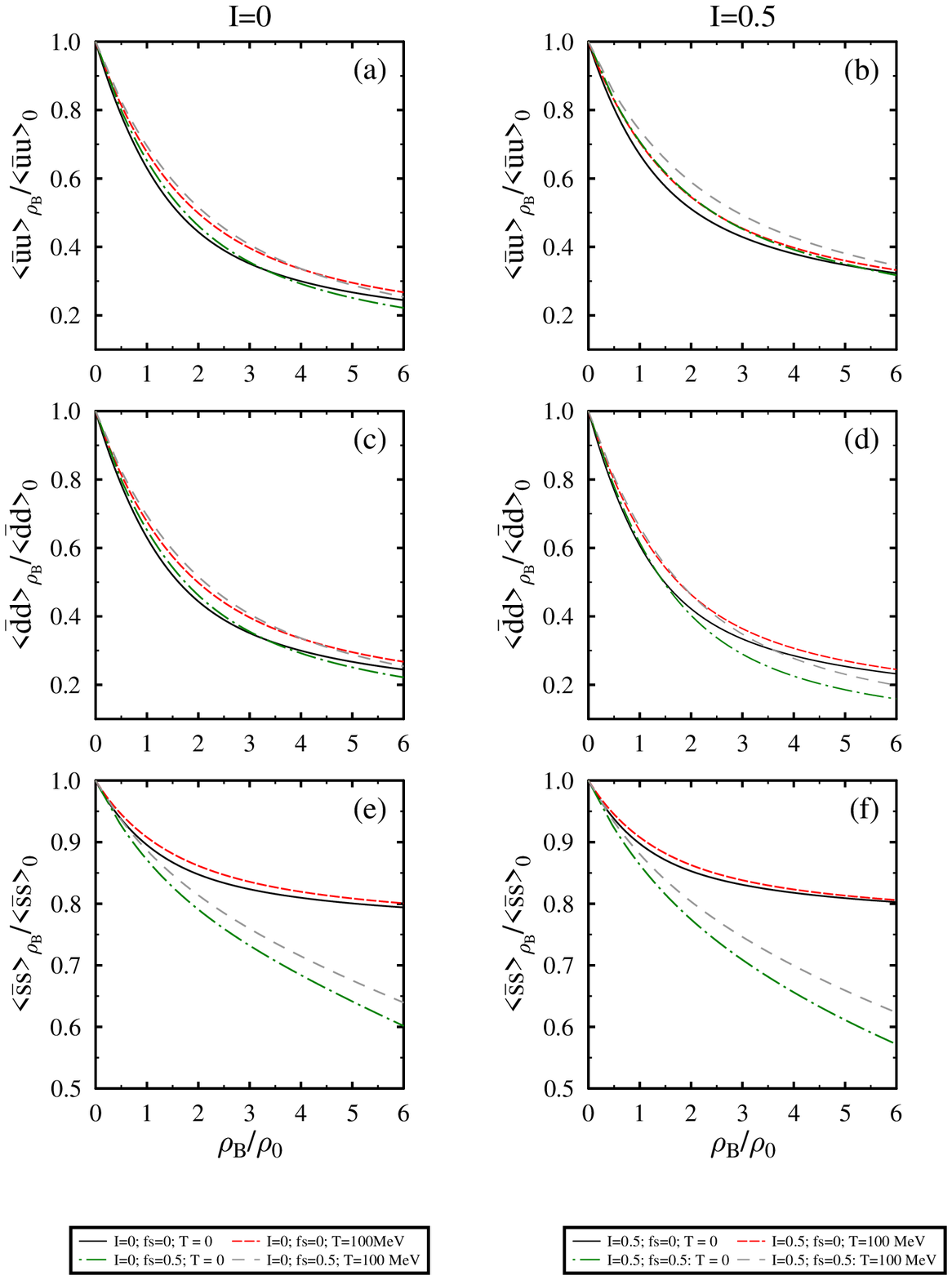}  
\caption{(Color online)
In the above figure, we plot the ratio of in-medium quark condensate to vacuum condensate,
 as a function of
 baryonic density, $ \rho_B$ (in units of nuclear saturation density, $\rho_0$). 
 We compare the results at isospin asymmetric parameters $I$ = 0 and 0.5. 
 For each value of isospin asymmetry parameter $I$, the results are
 shown for 
 strangeness fractions $f_s$ =0 and 0.5
 and temperatures $T$ = 0 and 100 MeV.}
\label{lu}
\end{figure}

\begin{figure}
\includegraphics[width=16cm,height=10cm]{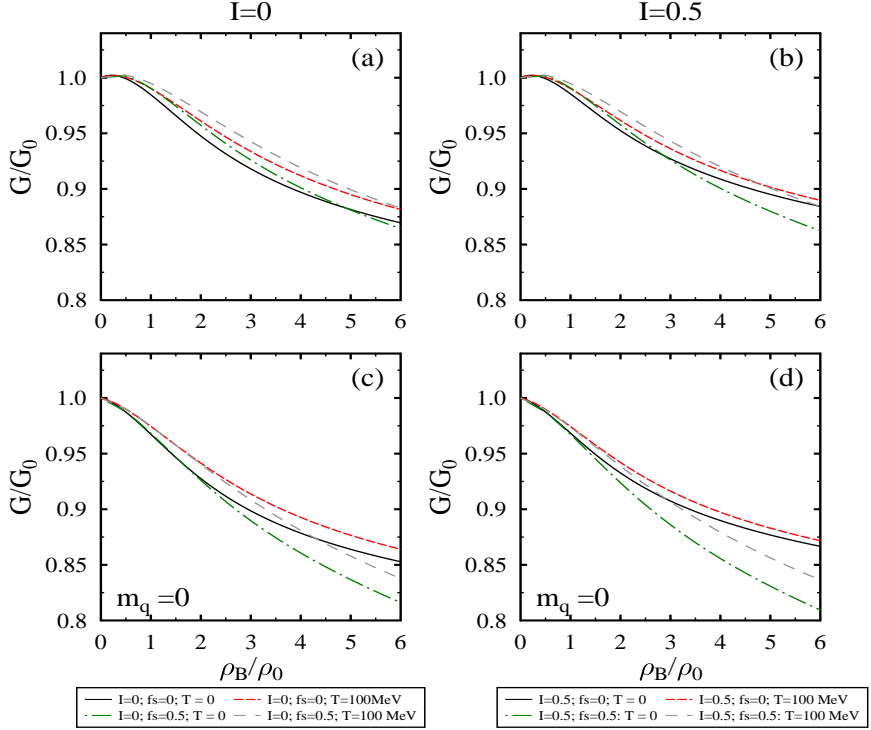}  
\caption{(Color online)
In the above figure, we plot the ratio of in-medium
scalar gluon condensate to the vacuum value of gluon condensate, 
 as a function of 
 baryonic density, $ \rho_B$ (in units of nuclear saturation density, $\rho_0$).  
 We compare the results at isospin asymmetric parameters $I$ = 0 and 0.5. 
 For each value of isospin asymmetry parameter $I$, the results are
 shown for 
 strangeness fractions $f_s$ =0 and 0.5
 and temperatures $T$ = 0 and 100 MeV. 
  In y-axis, $G$ = $\left\langle \frac{\alpha_s}{\pi}G_{\mu\nu}^{a} G^{\mu\nu a}\right\rangle $ and $G_0$ denotes the expectation value of gluon condensate
  for zero density and temperature.}
\label{gl2}
\end{figure}

In 
Figs. \ref{lu} and \ref{gl2},
the ratios of,  in-medium value 
 to the vacuum value of 
scalar quark and gluon condensates respectively, are
plotted
 as a function of baryonic density of hadronic medium.
We show the results for light quark condensates, 
$\left\langle \bar{u}u \right\rangle$, $\left\langle \bar{d}d \right\rangle$, the strange quark condensate,
$\left\langle \bar{s}s \right\rangle$
and the gluon condensates,
$\left\langle \frac{\alpha_s}{\pi}G_{\mu\nu}^{a} G^{\mu\nu a}\right\rangle $. 
In Table \ref{table_condensatess}, 
we tabulate the values of
the ratio of these condensates for quantitative
understanding. 
In this table, $\left\langle \bar{u}u\right\rangle _{0}$,
 $\left\langle \bar{d}d\right\rangle _{0}$ and
  $\left\langle \bar{s}s\right\rangle _{0}$,
  denotes the vacuum values of quark condensates and
in chiral SU(3) model these are, $-1.401 \times 10^{-2}$ GeV$^{3}$,
$-1.401 \times 10^{-2}$ GeV$^{3}$ and $-4.671 \times 10^{-2}$ GeV$^{3}$ respectively.
We observe that, for a given value 
of temperature $T$,
isospin asymmetry parameter $I$,
and the strangeness fraction $f_{s}$, as a function of 
baryonic density, the
magnitude of the values of quark condensate decreases w.r.t. vacuum value.
This is because, the values of quark condensates
are proportional to the scalar fields $\sigma$ and
$\zeta$ (see Eqs. (\ref{qu}) to (\ref{qs})). As discussed above, the magnitude of these scalar fields undergo drop
as a function of density of baryonic matter
and this further cause a decrease in the magnitude of quark condensates.
As we move from symmetric to the asymmetric medium,
due opposite contribution of scalar-isovector
mesons $\delta$,
 the values of
condensate $\left\langle \bar{u}u \right\rangle$
 increases, whereas that of $\left\langle \bar{d}d\right\rangle$
decreases.

In isospin symmetric medium ($I$ = 0), below baryon density 
$\rho_B$ = $3.3 \rho_0$, as we move from 
non-strange medium, i.e., $f_s$ = 0, to strange medium with 
$f_s$ = 0.5, the magnitude of light quark condensates $\left\langle \bar{u}u\right\rangle $
increases. 
Above  baryon density, 
$\rho_B$ = 3.3 $\rho_0$, the magnitude of the 
values of light quark condensates is more in 
non strange medium ($f_s$ = 0) as compared 
to strange medium with  finite $f_s$.
  From Fig. \ref{lu}, we observe that, as a function of
  density of baryonic matter, we always observe a decrease 
  in the values of quark condensates.
  This support the expectation of chiral symmetry 
  restoration at high baryonic density.
  However, in linear Walecka model \cite{quark4,quark5} and also in Dirac-Brueckner
  approach \cite{quark3}, the values of quark condensates are
  observed to increase at higher values of baryonic
  density and causes hindrance to 
  chiral symmetry restoration. The possible reason for this may be 
  that the chiral invariance property was not considered in
  these calculations \cite{quark3}.
   As the strange quark condensates $\left\langle \bar{s}s\right\rangle $,
  is proportional to the
  strange scalar-isoscalar field $\zeta$, 
  therefore, the behavior of this field as
  a function of various parameters
  of the medium is also reflected in the
  values of strange quark condensate $\left\langle \bar{s}s\right\rangle $.
  For fixed baryon density, $\rho_B$ and isospin asymmetry
  parameter $I$, as we move from 
  non-strange to strange hadronic medium,
  the values of strange condensate  
  decreases.

\begin{table}
\begin{tabular}{|l|l|l|l|l|l|l|l|l|l|}
\hline
 & & \multicolumn{4}{c|}{I=0}    & \multicolumn{4}{c|}{I=0.5}   \\
\cline{3-10}
&$f_s$ & \multicolumn{2}{c|}{T=0} & \multicolumn{2}{c|}{T=100}& \multicolumn{2}{c|}{T=0}& \multicolumn{2}{c|}{T=100}\\
\cline{3-10}
&  &$\rho_0$&$4\rho_0$ &$\rho_0$  &$4\rho_0$ & $\rho_0$ &$4\rho_0$&$\rho_0$&$4\rho_0$ \\ \hline

$\frac{<u \bar{u}> }{<u \bar{u}>_0}$& 0& 0.62&0.29 &0.67&0.33&0.66&0.38&0.70 &0.39\\ \cline{2-10}

&0.5&0.65  &0.29  & 0.69 & 0.33  & 0.70 &0.39 & 0.74 & 0.42 \\ \cline{1-10}
$\frac{<d\bar{d}>}{<d\bar{d}>_0}$&0&0.63 &0.29  & 0.67 & 0.35 & 0.60 & 0.28 & 0.65 &0.30 \\   \cline{2-10}

&0.5&0.65 &0.29&0.69&0.33&0.61&0.22 &0.66 &0.27\\ \hline

$\frac{<s\bar{s}>}{<s\bar{s}>_0}$&0&0.99 & 0.80 & 0.90 & 0.81 & 0.89 & 0.81 & 0.90 &0.82 \\  \cline{2-10}

&0.5&0.87 & 0.68 &0.88 & 0.71 & 0.86 & 0.65 & 0.88& 0.69 \\  \hline

$\frac{G}{G_0}$&0 &0.98 &0.89 &0.99&0.91 &0.98&0.90&0.99& 0.91\\  \cline{2-10}

& 0.5&0.99 & 0.90&0.99& 0.91&0.99&0.90&0.99&0.92\\ \hline
$\frac{G}{G_0}$ $(m_q = 0)$&0 &0.96 &0.87 &0.97&0.89 &0.96&0.88&0.97& 0.89\\  \cline{2-10}

& 0.5&0.96 & 0.86&0.97& 0.88&0.96&0.85&0.97&0.87\\ \hline

\end{tabular}
\caption{In the above table,  we tabulate the ratio of in-medium value to the vacuum
value of scalar quark and gluon condensates.}
\label{table_condensatess}
\end{table}
 In Figs. \ref{gl2}(a) and
 \ref{gl2}(b), 
  we consider the
  contribution of finite quark mass term in the
  calculation of gluon condensates, 
  whereas Figs. \ref{gl2}(c) and \ref{gl2}(d)
  are
  without the effect of quark mass term.
  We observe that, as a function of baryonic density of the
  hadronic medium, the values of gluon condensates decreases.
 As one can see from Figs. \ref{gl2}(c) and 
 \ref{gl2}(d), the effect of
 strangeness fractions are more visible at
 higher baryon densities.
The calculations show that the
gluon condensates have small density dependence as compared to quark condensates.
This observation is consistent with earlier model independent
calculations by Cohen \cite{cohen} and also with the QMC model calculation
in Ref. \cite{quark6}.

Also it is observed that, at zero baryonic density, 
the behavior of scalar fields and hence the scalar quarks and gluon condensates,
as a function of temperature of the  medium is opposite to 
the situation of finite baryonic density.
At $\rho_B = 0$,
the drop in the scalar fields and scalar condensates increases (magnitude decreases) with increase in temperature of the medium. 
Recall from previous discussion, at finite $\rho_B$,
the drop in the scalar fields and hence scalar condensates was decreasing (magnitude 
increasing) with increase in the temperature of the medium (see Figs. \ref{lu} and \ref{gl2}).
However, for zero baryonic density, the change in the magnitude of scalar fields
with the temperature is very small in the hadronic medium.
For example,
at $\rho_B = 0$, the magnitude of the
scalar field $\sigma$ ($\zeta$)
decreases by $0.06\%$ ($0.02\%$) as one move from $T = 0$ to $T$ = 100 MeV.
For a change of temperature from $T = 100$ to $T = 150$
MeV, the magnitude of $\sigma$ ($\zeta$) decreases by $2.5\%$ ($0.8\%$).
The values of, light quark condensate, strange quark
condensate, and the
gluon condensate with zero quark mass term, at $\rho_B$ = 0, changes
by $2.75\%$, $0.85\%$, and $0.08\%$ respectively, as $T$ is changed from 0 to 150 MeV.
However, at very high temperature, possibly above critical temperature, the 
values of scalar condensates may change sharply \cite{josef}.
As we discussed earlier, in literature
the values of scalar condensates at finite temperatures and zero baryonic
densities are calculated using the pion bath contributions \cite{kwon1}.
Our observations on the behavior of condensates
as a function of temperature, for zero baryonic density, are
in accordance with  the earlier findings \cite{condt1,condt2,condt3}.

\subsection{Heavy Vector Mesons ($D^{*}$, $B^{*}$,
 $D_s^{*}$ and $B_s^{*}$):}
\begin{figure}
\includegraphics[width=20cm,height=20cm]{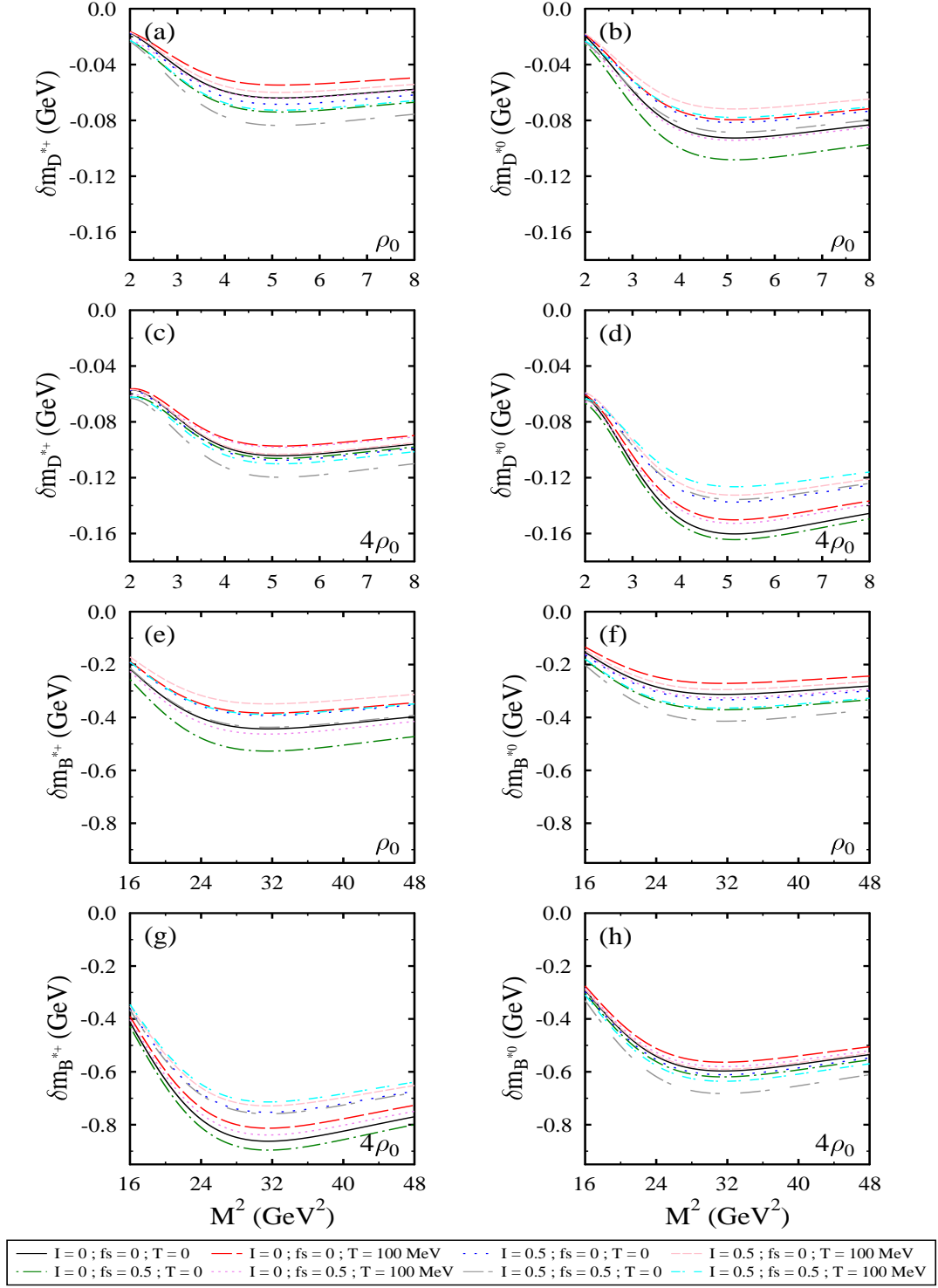}  
\caption{(Color online)
In the above figure,
 the variation of mass shift  of
 vector mesons, $D^{*}$ ($D^{*+}$ and $D^{*0}$)
  and $B^{*}$ ($B^{*+}$ and $B^{*0}$),
   is shown as a function 
 of squared Borel mass parameter $M^2$. 
We compare the results at 
temperatures $T$ = 0 and $T = 100$ MeV. 
 For each value of temperature, the results are
 shown for isospin asymmetric parameters $I$ = 0 and 0.5 and the
 strangeness fractions $f_s$ = 0 and 0.5.}
\label{dstbstmass}
\end{figure}

\begin{figure}
\includegraphics[width=20cm,height=20cm]{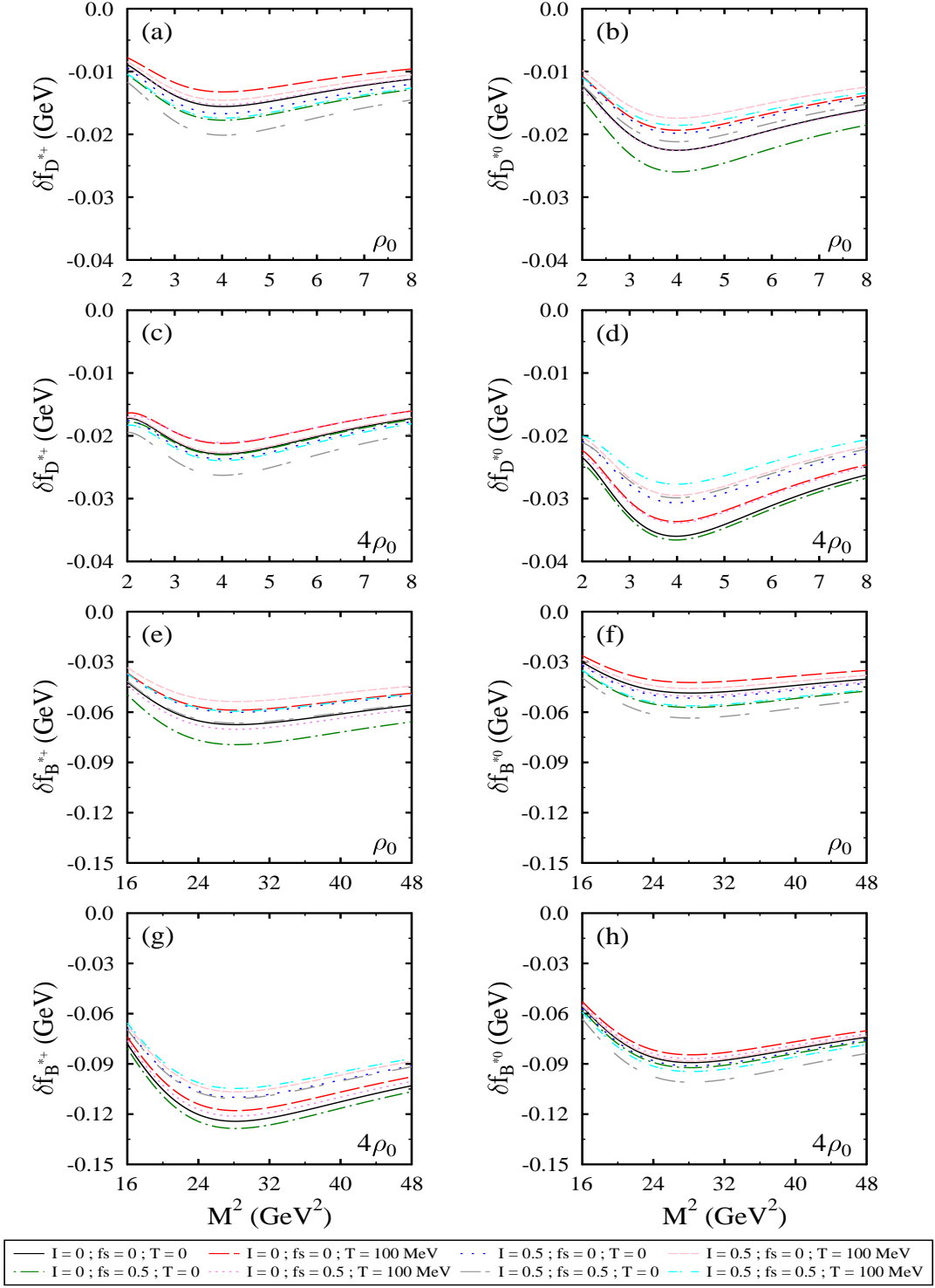}  
\caption{(Color online)
In the above figure,
 the variation of shift in decay constant of
 vector mesons $D^{*}$ ($D^{*+}$ and $D^{*0}$)
  and $B^{*}$ ($B^{*+}$ and $B^{*0}$),
   is shown as a function 
 of squared Borel mass parameter $M^2$. 
We compare the results at 
temperatures $T$ = 0 and $T = 100$ MeV. 
 For each value of temperature, the results are
 shown for isospin asymmetric parameters $I$ = 0 and 0.5, and the
 strangeness fractions $f_s$ = 0 and 0.5.}
\label{dstbstdecay}
\end{figure}

 In Fig. \ref{dstbstmass}
  (Fig. \ref{dstbstdecay}), 
 we show the 
 variation of mass shift (shift in decay constant) of
   vector mesons  $D^{*}$ and $B^{*}$,  
  as a function of squared Borel mass parameter $M^2$.
  In each subplot, we  compare the results 
  for temperatures $T$ = 0 and $T$ = 100 MeV.
For each value of temperature,
the results are plotted for isospin asymmetry parameters
    $I$ = 0 ($f_s$ = 0 and 0.5) and $I$ = 0.5
  ($f_s$ = 0 and 0.5).
   We present the results at baryonic
  densities $\rho_0$ and $4\rho_0$.
  In Table \ref{dstbstmasstable} 
  (Table \ref{dstbstdeacytable}), we have written the values of
  shift in mass (decay constant) of these vector mesons.
\begin{table}
\begin{tabular}{|l|l|l|l|l|l|l|l|l|l|}
\hline
 & & \multicolumn{4}{c|}{I=0}    & \multicolumn{4}{c|}{I=0.5}   \\
\cline{3-10}
&$f_s$ & \multicolumn{2}{c|}{T=0} & \multicolumn{2}{c|}{T=100}& \multicolumn{2}{c|}{T=0}& \multicolumn{2}{c|}{T=100}\\
\cline{3-10}
&  &$\rho_0$&$4\rho_0$ &$\rho_0$  &$4\rho_0$ & $\rho_0$ &$4\rho_0$&$\rho_0$&$4\rho_0$ \\ \hline

$\delta m_{D^{*+}}$ & 0& -64&-104 &-54&-97&-68&-104&-60 &-97\\ \cline{2-10}
&0.5&-74  &-106  &-64 & -98 &-83 &-119 & -72 & -110 \\ \cline{1-10}

$\delta m_{D^{*0}}$&0&-92 & -160 & -79 &-150 &-81 &-132 & -71 &-132 \\  \cline{2-10}
&0.5&-108 &-164 &-94 &-152 & -88 &-135 & -77&-126 \\  \hline

$\delta m_{B^{*+}}$& 0&-443&-862 &-383&-813&-392&-752&-348 &-728\\ \cline{2-10}
&0.5&-527  &-896  &-462 & -838 &-437 &-758 & -388 & -713 \\ \cline{1-10}
$\delta m_{B^{*0}}$&0&-312 & -596 & -271 &-563 &-333 &-610 & -294 &-590 \\  \cline{2-10}
&0.5&-370 &-618 &-326 &-580 & -414 &-682 & -364&-635 \\  \hline
\end{tabular}
\caption{In the above table, we tabulate the values of mass shift (in units of MeV) of
heavy vector mesons $D^{*}$ ($D^{*+}$ and $D^{*0}$) and  $B^{*}$ ($B^{*+}$ and $B^{*0}$).}
\label{dstbstmasstable}
\end{table}
       The difference in the masses of
         $D^{*+}$ ($B^{*+}$) and $D^{*0}$ ($B^{*0}$)
       mesons in the symmetric nuclear medium is due to
       the different masses of $u$ and $d$ quarks,
       considered in the present investigation.
Also we observe that, 
for a given value of baryonic density and 
strangeness fraction,
both, the masses and decay constants, of
$D^{*+}$ and $B^{*0}$ ($D^{*0}$ and $B^{*+}$) mesons, undergo more (less)  drop in
asymmetric medium as compared to symmetric medium.
Note that, the $D^{*+}$ and $B^{*0}$ mesons contain the
light $d$ quark, whereas the $D^{*0}$ and $B^{*+}$
mesons have light $u$ quark.
As discussed earlier, the behavior of $\left\langle\bar{d}d \right\rangle$
and $ \left\langle \bar{u}u\right\rangle$ condensates is
opposite as a function of asymmetry of
the medium and this causes
the observed behavior of  
 $D^{*+}$ ($B^{*0}$)
and $D^{*0}$ ($B^{*+}$) mesons, as a function of asymmetry of the
medium.
As compared to nuclear medium ($f_s = 0$),
 the drop in the masses and decay constant of $D^{*}$ and $B^{*}$
mesons is more in strange hadronic medium (finite $f_{s}$).  
 Also, the drop in the  masses of $D^{*}$ and $B^{*}$ mesons
 increases with increase in the baryonic density of the medium.
 At given finite baryonic density, the non-zero temperature of the medium cause less drop
 in the masses as well as decay constant of $D^{*}$ and
 $B^{*}$ mesons, as compared to zero temperature situation. 
 This is because of increase in the values of scalar fields
 and condensates as a function of temperature of the medium, at  given
 finite baryonic density \cite{quarkmean1}.
  
\begin{table}
\begin{tabular}{|l|l|l|l|l|l|l|l|l|l|}
\hline
 & & \multicolumn{4}{c|}{I=0}    & \multicolumn{4}{c|}{I=0.5}   \\
\cline{3-10}
&$f_s$ & \multicolumn{2}{c|}{T=0} & \multicolumn{2}{c|}{T=100}& \multicolumn{2}{c|}{T=0}& \multicolumn{2}{c|}{T=100}\\
\cline{3-10}
&  &$\rho_0$&$4\rho_0$ &$\rho_0$  &$4\rho_0$ & $\rho_0$ &$4\rho_0$&$\rho_0$&$4\rho_0$ \\ \hline

$\delta f_{D^{*+}}$&0&-15 &-23  &-13 &-21 &-16 & -23 &-14 &-22 \\   \cline{2-10}
&0.5&-17 &-23&-15&-21&-20&-26 &-17 &-23\\ \hline

$\delta f_{D^{*0}}$&0 &-22 &-36 &-19&-33 &-19&-30&-17&-29\\  \cline{2-10}
& 0.5&-25 &-36&-22&-33&-21&-29&-18&-27\\ \hline

$\delta f_{B^{*+}}$&0&-67 &-124  &-58 &-117 &-60 & -109 &-53 &-106 \\   \cline{2-10}
&0.5&-79 &-128&-70&-121&-66&-110 &-59 &-104\\ \hline

$\delta f_{B^{*0}}$&0 &-48 &-89 &-42&-84 &-51&-91&-45&-88\\  \cline{2-10}
& 0.5&-57 &-92&-50&-86&-63&-100&-56&-94\\ \hline

\end{tabular}
\caption{In the above table, we tabulate the values of  shift in decay constant (in units of MeV) of
heavy vector meson $D^{*}$ ($D^{*+}$ and $D^{*0}$) and  $B^{*}$ ($B^{*+}$ and $B^{*0}$).}
\label{dstbstdeacytable}
\end{table}
 
\begin{figure}
\includegraphics[width=20cm,height=20cm]{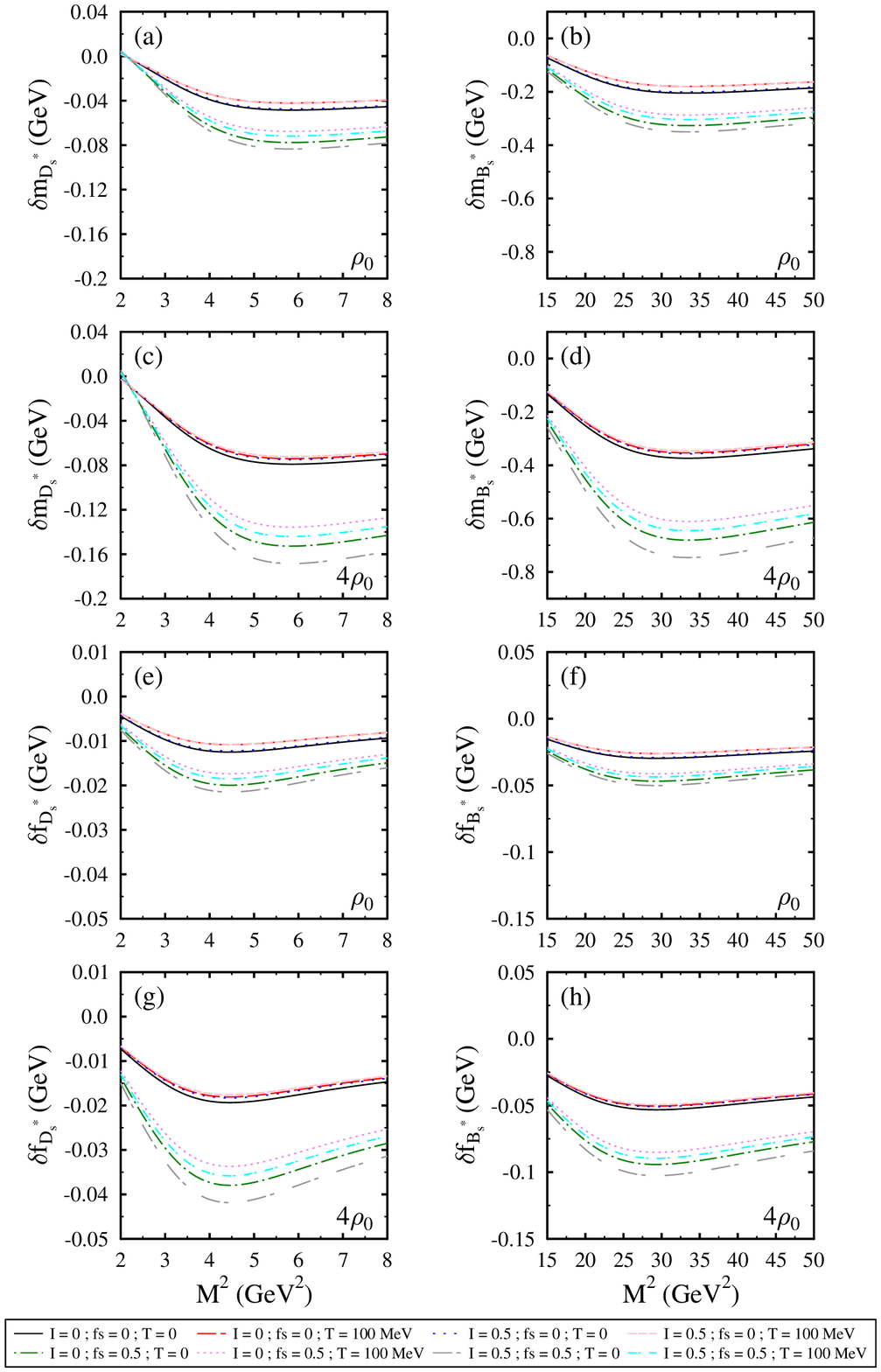}  
\caption{(Color online)
In the above figure, the variation of shift in mass and 
decay constant of
 strange vector mesons $D_{s}^{*}$ and $B_{s}^{*}$, is shown as a function 
 of squared Borel mass parameter $M^2$. 
 We show the results at temperatures $T = 0$ and $T = 100$ MeV.
 For each value of temperature, the
 results are shown for isospin asymmetry parameter, $I$ =
 0 and 0.5
 and the 
 strangeness fractions, $f_s$ = 0 and 0.5.}
\label{DsBsmassdecay}
\end{figure}

\begin{table}
\begin{tabular}{|l|l|l|l|l|l|l|l|l|l|}
\hline
 & & \multicolumn{4}{c|}{I=0}    & \multicolumn{4}{c|}{I=0.5}   \\
\cline{3-10}
&$f_s$ & \multicolumn{2}{c|}{T=0} & \multicolumn{2}{c|}{T=100}& \multicolumn{2}{c|}{T=0}& \multicolumn{2}{c|}{T=100}\\
\cline{3-10}
&  &$\rho_0$&$4\rho_0$ &$\rho_0$  &$4\rho_0$ & $\rho_0$ &$4\rho_0$&$\rho_0$&$4\rho_0$ \\ \hline

$\delta m_{D_s^{*}}$& 0&-48&-79 &-42&-74&-47&-74&-42 &-72\\ \cline{2-10}
&0.5&-77  &-152  &-67 & -135 &-83 &-168 & -71 & -144 \\ \cline{1-10}

$\delta m_{B_s^{*}}$&0&-204 & -373 & -179 &-353 &-200 &-357 & -174 &-345 \\  \cline{2-10}
&0.5&-326 &-680 &-287 &-610 & -350 &-745 & -304&-645 \\  \hline

$\delta f_{D_s^{*}}$&0&-12 &-19  &-10 &-18 &-12 & -18 &-10&-17\\   \cline{2-10}
&0.5&-19 &-37&-17&-33&-21&-41 &-18 &-35\\ \hline

$\delta f_{B_s^{*}}$&0 &-29 &-53 &-26&-50 &-29&-51&-26&-49\\  \cline{2-10}
& 0.5&-46 &-94&-41&-85&-50&-102&-43&-89\\ \hline

\end{tabular}
\caption{In the above table, we tabulate the values of  shift in masses
and decay constants (in units of MeV) of
heavy strange vector mesons $D_s^{*}$ and  $B_s^{*}$.}
\label{dsstbsstmassdecaytable}
\end{table}
 
In Fig. \ref{DsBsmassdecay} and Table \ref{dsstbsstmassdecaytable},
  we present the
variations of mass shift and shift in decay constant of $D_s^*$
 and $B_s^*$ mesons.
 The charmed strange and bottom strange mesons have one strange quark $s$
 and one heavy quark.
 As discussed earlier, the in-medium properties of these mesons are calculated through the presence of strange quark
 condensate $\left\langle \bar{s}s\right\rangle $, in OPE
 side of QCD sum rule equations \cite{wang2}.
   The negative values of shift in masses and decay constants
   show that these parameters of
   strange charmed and bottom vector
   meson undergo drop in the hadronic medium.
 We notice that,  the effect of increasing either the
  density or strangeness fraction of the
  medium is to decrease 
    the masses and decay constants of  $D_s^*$  and $B_s^*$ mesons.
    However, as observed for  $D^*$  and $B^*$ mesons,
    the masses and decay constants of
     $D_s^*$  and $B_s^*$ also increase with increase in temperature
     of the medium.

It may be noted that, among the various condensates
present in the QCD sum rule equations, the scalar quark condensates $\left\langle \bar{q}q\right\rangle $, have
 largest contribution for the medium modification of $D$ and $B$ meson properties.
 For example, if
 all the condensates are set to zero, except 
$<\bar{q}q>$, then shift in mass (decay constant) 
for $D^{*+}$ meson in  symmetric nuclear
medium ($f_s$ = 0 and $I$ = 0) is observed to
 be -69(-17.8) MeV for $\rho_B$ = $\rho_0$
 and can be compared to the values   -64 (-15) MeV
 evaluated in the presence of all condensates.
 Also, the condensate $<q^\dag i D_0 q>$ is not
evaluated within chiral SU(3) model. 
In literature, the condensate 
$\langle q^\dag i D_0q\rangle$
is available in terms of
linear density formula only and hence same is used in our calculations.
Dependence of in-medium properties of $D$ mesons on this condensate
is negligible, at least upto linear density approximation, 
 as was also found in literature \cite{haya1,wang1,wang2,wang3}. 
 However,  
at high density it might become important. 
Neglecting $<q^\dag i D_0 q>$ only, the values of mass shift (decay shift)
in symmetric medium, at $\rho_B$ = $\rho_0$, $2\rho_0$ and 
$4\rho_0$
are
 observed to be -62(-15.5), -89(-21) and -97(-21) MeV, respectively.
 These values can be compared with
 the values,  
 -64(-15), -92(-22) and -104(-23) MeV, respectively,
 which are computed in the situation when 
 we considered the contribution
of  $<q^\dag i D_0 q>$ within linear density approximation.
Thus, the mass shift (decay shift) for $D^{\star +}$ mesons
at density $2\rho_0$ is changed  by 
$3.58\% (2.3\%)$,
whereas, at $4\rho_0$, the
percentage change is $6.7\% (4.8\%)$.

The observed modifications of
 masses and decay constant of vector mesons are
 not much sensitive to the 
 values of coupling constant
 of heavy baryons with nucleons
 and charmed/bottom mesons. 
 For example,
 in symmetric nuclear matter, at nuclear saturation density  $\rho_0$,
  for an increase (decrease)
of $20 \%$ in the value of coupling constant, 
the shift in the mass of $D^{* +}$ mesons decreases (increases)  by
$1.4 \%$ ($2.8 \%$) only.
As we discussed earlier, the continuum threshold parameters are chosen so as
to reproduce  the experimental observed vacuum values of
the masses of charmed and bottom mesons mesons. 
The values of threshold parameter effect significantly the
masses and decay constant of $D$ and $B$ mesons.
For example, for a decrease (increase) of $10 \%$ in the
values of  threshold parameter $s_0$,
in symmetric nuclear matter, at $\rho_B$ = $\rho_0$,
 an increase (decrease) of $29 \%$ ($15 \%$)  is observed in the
 value of  mass shift of 
 $D^{* +}$ mesons. 

 \subsection{Heavy Axial-Vector Mesons ($D_{1}$, $B_{1}$,
 $D_{1s}$ and $B_{1s}$):}
\begin{figure}
\includegraphics[width=20cm,height=20cm]{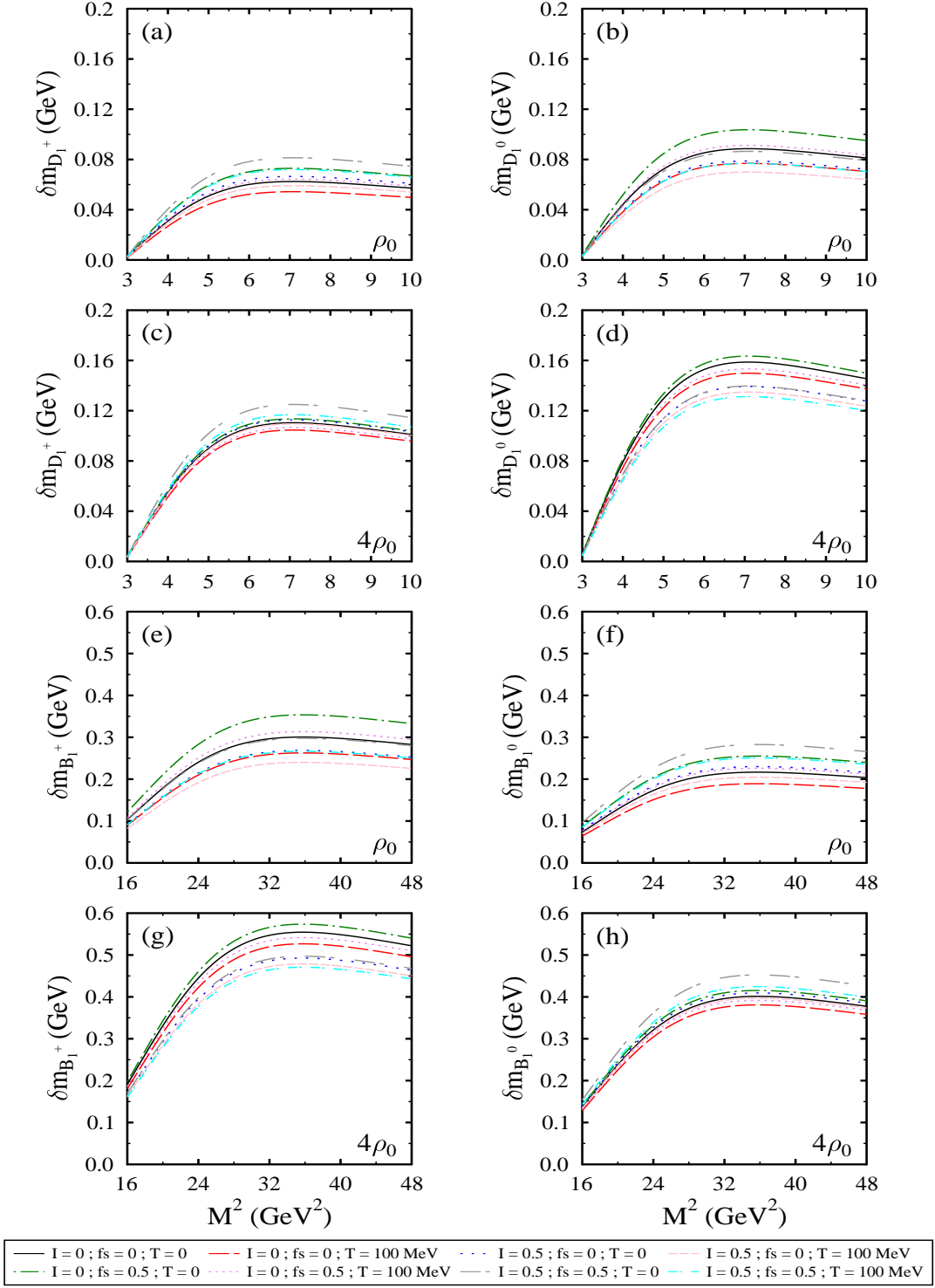}  
\caption{(Color online)
In the above figure,
 the variation of mass shift  of
 axial-vector mesons, $D_{1}$ ($D_{1}^{+}$ and $D_{1}^{0}$)
  and $B_{1}$ ($B_{1}^{+}$ and $B_{1}^{0}$),
   is shown as a function 
 of squared Borel mass parameter $M^2$. 
We compare the results at 
temperatures $T$ = 0 and $T = 100$ MeV. 
 For each value of temperature, the results are
 shown for isospin asymmetric parameters $I$ = 0 and 0.5 and the
 strangeness fractions $f_s$ = 0 and 0.5.}
\label{D1B1mass}
\end{figure}

\begin{figure}
\includegraphics[width=20cm,height=20cm]{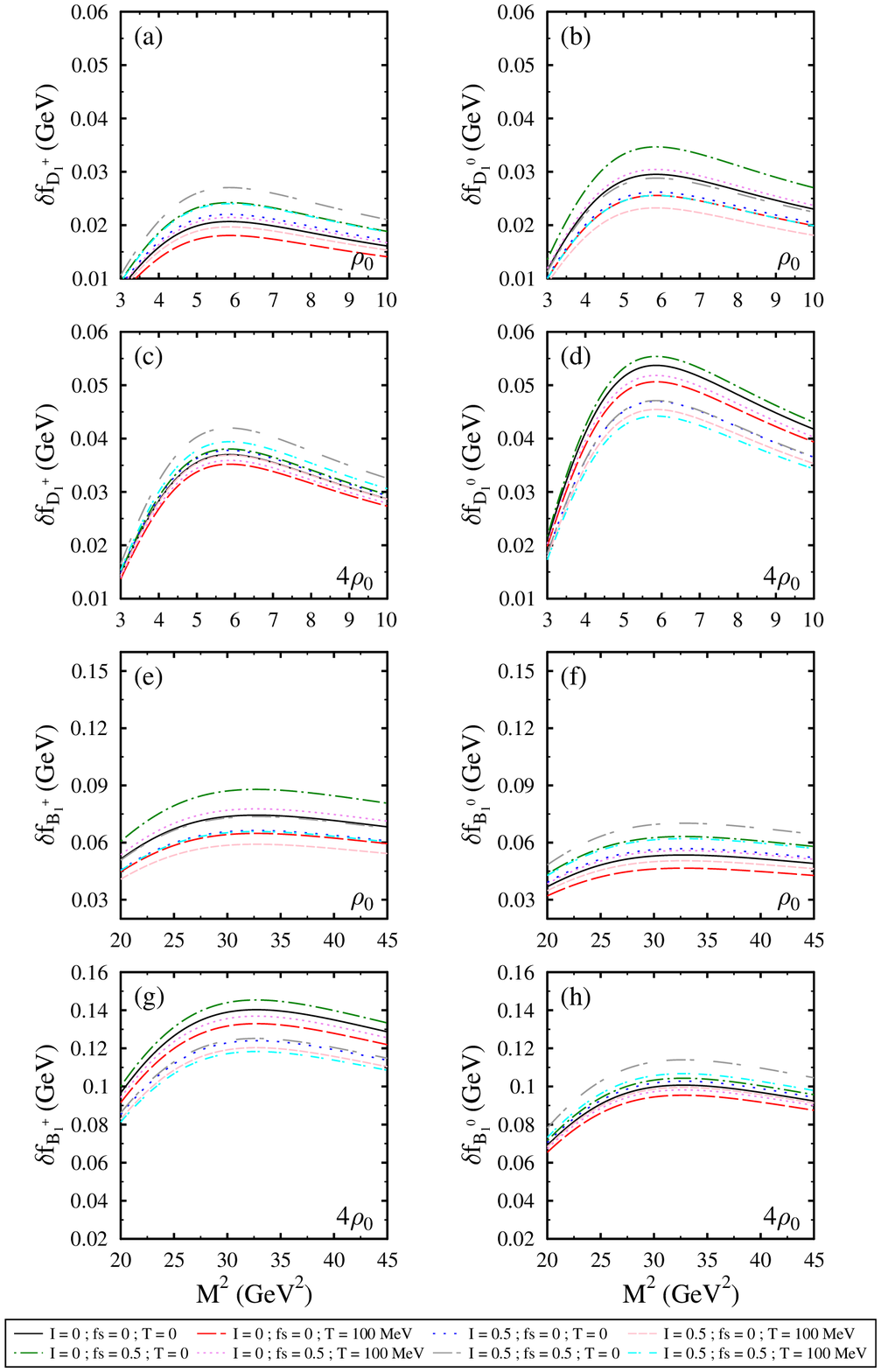}  
\caption{(Color online)
In the above figure,
 the variation of  shift in decay constant  of
axial-vector mesons, $D_{1}$ ($D_{1}^{+}$ and $D_{1}^{0}$)
  and $B_{1}$ ($B_{1}^{+}$ and $B_{1}^{0}$),
   is shown as a function 
 of squared Borel mass parameter $M^2$. 
We compare the results at 
temperatures $T$ = 0 and $T = 100$ MeV. 
 For each value of temperature, the results are
 shown for isospin asymmetric parameters $I$ = 0 and 0.5 and the
 strangeness fractions $f_s$ = 0 and 0.5.}
\label{D1B1decay}
\end{figure}

\begin{figure}
\includegraphics[width=20cm,height=20cm]{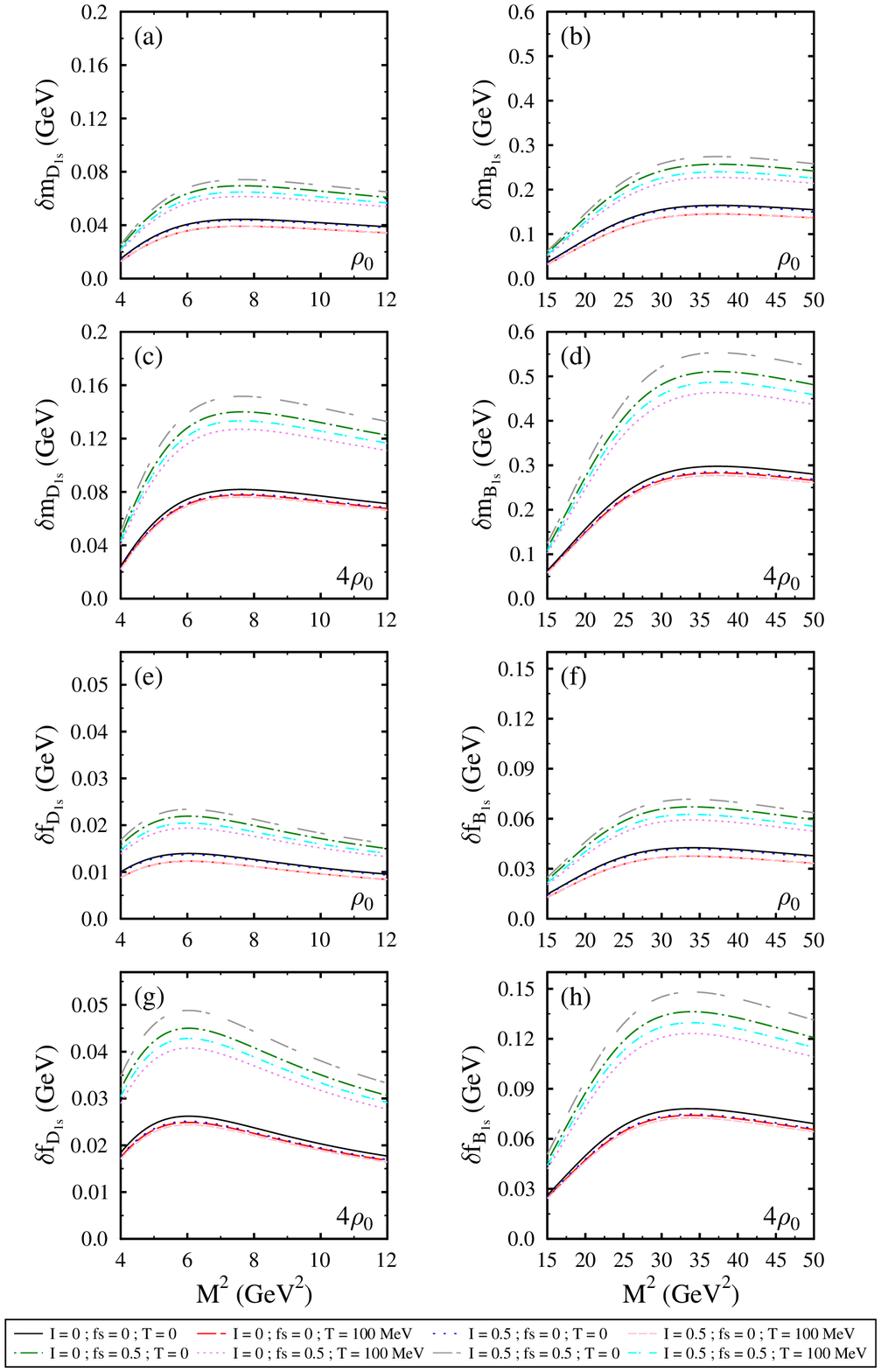}  
\caption{(Color online)
In the above figure, the variation of shift in mass and 
decay constant of
 strange axial-vector mesons $D_{1s}$ and $B_{1s}$,
 is shown as a function 
 of squared Borel mass parameter $M^2$. 
 We show the results at temperatures $T = 0$ and $T = 100$ MeV.
 For each value of temperature, the
 results are shown for isospin asymmetry parameter, $I$ =
 0 and 0.5
 and the 
 strangeness fractions, $f_s$ = 0 and 0.5.}
\label{Ds1Bs1massdecay}
\end{figure}

Figures \ref{D1B1mass} and \ref{D1B1decay} show the variation 
of mass shift  and shift in decay constant, respectively,
of axial-vector mesons [$D_1$ ($D_1^{0}$ and $D_1^{+}$)
and $B_1$ ($B_1^{0}$ and $B_1^{+}$)],
as a function of squared Borel mass parameter, i.e., $M^2$. 
  In Tables \ref{D1B1masstable} and \ref{D1B1decaytable},
we tabulate the   
   values of shift in mass and decay constants of  
   these axial-vector mesons.
   Similarly, Fig. \ref{Ds1Bs1massdecay}
 and  Table \ref{D1sB1smassdecayshifttable},
are for the strange axial-vector, $D_{1s}$ and $B_{1s}$ mesons.   
 For a constant value of strangeness fraction $f_s$,
      and isospin asymmetric parameter $I$, as
      a function of baryonic density, 
      a positive shift in masses and decay constants
        of axial-vector  
         mesons was
        observed.     
     For a given density and isospin asymmetry, the
      finite strangeness fraction of the medium also causes an 
      increase in the masses and decay constants 
      of above axial-vector mesons.
  For the axial-vector $D_1$ and $B_1$ meson
   doublet, as a function of isospin asymmetry of the medium, 
       the values of mass shift and decay shift of $D_1^{0}$ ($B_1^{0}$) 
       meson decreases (increases),  whereas that
       of  $D_1^{+}$ ($B_1^{+}$)  increases (decreases).
       As discussed earlier, in the case of vector mesons,
       the reason for opposite behavior of $D$ and $B$ mesons
       as a function of isospin asymmetry of medium is the
       presence of light $u$ quark  in 
       $D_1^{0}$ and $B_1^{+}$ mesons, whereas  
       the mesons $D_1^{+}$ and $B_1^{0}$ have light $d$ quark.
       At finite baryonic density, for nuclear as
       well as  strange hadronic medium, the effect
        of finite temperature is to cause an increase in the values of positive mass
       shift of the axial-vector mesons w.r.t. zero temperature case.
       

\begin{table}
\begin{tabular}{|l|l|l|l|l|l|l|l|l|l|}
\hline
 & & \multicolumn{4}{c|}{I=0}    & \multicolumn{4}{c|}{I=0.5}   \\
\cline{3-10}
&$f_s$ & \multicolumn{2}{c|}{T=0} & \multicolumn{2}{c|}{T=100}& \multicolumn{2}{c|}{T=0}& \multicolumn{2}{c|}{T=100}\\
\cline{3-10}
&  &$\rho_0$&$4\rho_0$ &$\rho_0$  &$4\rho_0$ & $\rho_0$ &$4\rho_0$&$\rho_0$&$4\rho_0$ \\ \hline

$\delta m_{D_1^{+}}$ & 0& 62&110 &54&104&66&112&59 &109\\ \cline{2-10}
&0.5&72  &113  &64 & 106 &81 &124 & 72 & 116 \\ \cline{1-10}

$\delta m_{D_1^{0}}$&0&88 &158 & 168 &149 &78 &140 &69 &134 \\  \cline{2-10}
&0.5&103 &163 &91 &153 & 86 &140 & 76&131 \\  \hline

$\delta m_{B_1^{+}}$& 0&300&554 &262&526&268&492&240 &478\\ \cline{2-10}
&0.5&353  &573  &313 &541 &297 &497 &266 &470 \\ \cline{1-10}
$\delta m_{B_1^{0}}$&0&216 & 401 &189 &380 &229 &409 &204 &396 \\  \cline{2-10}
&0.5&255 &415 &225 &391 &282 &452 &250&424 \\  \hline

\end{tabular}
\caption{In the above table, we tabulate the values of mass shift (in units of MeV) of
heavy axial-vector mesons $D_{1}$ ($D_{1}^{+}$ and $D_{1}^{0}$) and  $B_{1}$ ($B_{1}^{+}$ and $B_{1}^{0}$).}
\label{D1B1masstable}
\end{table}
\begin{table}
\begin{tabular}{|l|l|l|l|l|l|l|l|l|l|}
\hline
 & & \multicolumn{4}{c|}{I=0}    & \multicolumn{4}{c|}{I=0.5}   \\
\cline{3-10}
&$f_s$ & \multicolumn{2}{c|}{T=0} & \multicolumn{2}{c|}{T=100}& \multicolumn{2}{c|}{T=0}& \multicolumn{2}{c|}{T=100}\\
\cline{3-10}
&  &$\rho_0$&$4\rho_0$ &$\rho_0$  &$4\rho_0$ & $\rho_0$ &$4\rho_0$&$\rho_0$&$4\rho_0$ \\ \hline

$\delta f_{D_1^{+}}$&0&20 &36  &18 &35 &22 & 37 &19 &36 \\   \cline{2-10}
&0.5&24 &38&21&36&27&41 &24 &39\\ \hline

$\delta f_{D_1^{0}}$&0 &29 &53 &25&50 &26&46&23&45\\  \cline{2-10}
& 0.5&34 &55&30&51&28&47&25&44\\ \hline

$\delta f_{B_1^{+}}$&0&74 &140  &64 &132 &66 & 124 &59 &120 \\   \cline{2-10}
&0.5&87 &145&77&136&73&125 &65 &118\\ \hline

$\delta f_{B_1^{0}}$&0 &53&100 &46&95 &56&102&50&99\\  \cline{2-10}
& 0.5&63 &104&55&98&70&113&62&106\\ \hline

\end{tabular}
\caption{In the above table, we tabulate the values of  shift in decay constant (in units of MeV) of
heavy axial-vector mesons $D_{1}$ ($D_{1}^{+}$ and $D_{1}^{0}$) and  $B_{1}$ ($B_{1}^{+}$ and $B_{1}^{0}$).}
\label{D1B1decaytable}
\end{table}

\begin{table}
\begin{tabular}{|l|l|l|l|l|l|l|l|l|l|}
\hline
 & & \multicolumn{4}{c|}{I=0}    & \multicolumn{4}{c|}{I=0.5}   \\
\cline{3-10}
&$f_s$ & \multicolumn{2}{c|}{T=0} & \multicolumn{2}{c|}{T=100}& \multicolumn{2}{c|}{T=0}& \multicolumn{2}{c|}{T=100}\\
\cline{3-10}
&  &$\rho_0$&$4\rho_0$ &$\rho_0$  &$4\rho_0$ & $\rho_0$ &$4\rho_0$&$\rho_0$&$4\rho_0$ \\ \hline

$\delta m_{D_{1s}}$& 0&44&81 &39&77&43&78&39 &76\\ \cline{2-10}
&0.5&69  &140  &61 & 127 &74 &151 & 64 &133 \\ \cline{1-10}

$\delta m_{B_{1s}}$&0&164 & 297 & 145 &282 &161 &285 &145 &276 \\  \cline{2-10}
&0.5&257 &510 &227 &463 & 274 &553 & 239&486 \\  \hline

$\delta f_{D_{1s}}$&0&14 &26  &12 &24 &13 &25 &12&24\\   \cline{2-10}
&0.5&21 &45&19&40&23&48 &20 &42\\ \hline

$\delta f_{B_{1s}}$&0 &42 &78 &37&74 &41&74&37&72\\  \cline{2-10}
& 0.5&67 &136&59&123&71&148&62&129\\ \hline

\end{tabular}
\caption{In the above table, we tabulate the values of  shift in masses
and decay constants (in units of MeV) of
heavy strange axial-vector mesons $D_{1s}$ and  $B_{1s}$.}
\label{D1sB1smassdecayshifttable}
\end{table}

The mass shift and shift in decay
constants of charmed and bottom vector and axial-vector
mesons have been investigated in past using QCD sum rules
in symmetric nuclear matter
  only \cite{wang2, wang3}  . The values of mass shift
  for $D^{*}$ and $B^{*}$ mesons, 
  in leading order (next to leading order) calculations, were 
    -70 (-102) and -340 (-687) MeV, respectively.
    For the axial-vector $D_{1}$ and $B_{1}$ mesons,
    the above values of mass shift changes
    to 66 (97) and 260 (522) MeV, respectively.
    The values of shift in decay constant for
    $D^{*}$ and $B^{*}$ mesons, in
    leading order (next to leading order),
     are found to be -18(-26) and -55(-111) MeV, whereas
    for $D_{1}$ and $B_{1}$ mesons 
    these values changes to 21(31)
     and  67 (134) MeV, respectively.
    We can compare the above values of
    mass shift (decay shift) to our results -63 (-20), -312 (-48.9), 62 (20)
     and 216 (53) MeV 
    for $D^{*}$, $B^{*}$, $D_{1}$ and $B_{1}$ mesons,
    evaluated using $m_u$ = $m_d$ = 7 MeV,
    in symmetric nuclear matter ($I$ = 0 and $f_s$ = 0).
    As discussed earlier, in our calculations,
   for strange mesons, we used the
   value of decay constant equal to 1.16 times
   the decay constant of non-strange mesons.
   However, as an example, if we use, $f_{{D_s}}$ = $1.19 f_{{D}}$, 
 then the values of mass shift, at $\rho_0$ ($I$ = 0, $f_s$ = 0),
  for vector and axial-vector
 mesons, are observed to be $-46$ and $44$ MeV, respectively
 and can be compared with the values $-48$ and $42$
 MeV, respectively, obtained using, $f_{{D_s}}$ = $1.16 f_{{D}}$. 
    
In \cite{higler1,higler2}, it was observed that beyond linear
density approximation there
occur strong splitting between the particle-antiparticle
mass. In these work, 
the correlation function was parametrized, using the
Lehmann
representation, into the form,
$Im \, \pi(\omega,0) = \Delta \pi(\omega) =  \pi F_{+} \delta\left(\omega - m_{+} \right) 
- \pi F_{-} \delta\left(\omega + m_{-} \right) $, such that,
$m_{\pm} = m \pm \Delta m$ and $F_{\pm} = F \pm \Delta F$.
Here, $m$ is mass center and $\Delta m$ is 
the mass splitting between particle-antiparticles.
This correlation function was further used in the
odd and even part of QCD sum rules. However, 
in our present approach, following the work of \cite{haya1,wang2},
the correlation function is divided into the vacuum part
and a medium dependent part and the average mass-shift
was obtained. In Ref. \cite{higler8}, the mass-splitting between vector and axial-vector mesons had been studied using the 
Weinberg sum rules.  
    
    In the self consistent coupled channel approach of Ref. \cite{tolosint91},
    the repulsive optical potential for $D^{*}$
     mesons had been reported which is complementary to
    our calculations within QCD sum rules.
   The observed negative values of mass shift for $D^{*}$  and $B^{*}$
mesons
in nuclear and strange hadronic matter, in our calculations, favor the
decay of higher charmonium and bottomonium states to $D^{*}\bar{D}^{*}$ and
$B^{*}\bar{B}^{*}$
pairs and hence may cause the quarkonium suppression.
However, the axial-vector meson undergo a positive mass
shift in nuclear and strange hadronic medium and 
hence the possibility  of  decay of excited charmonium and bottomonium
states to $D_{1}\bar{D}_{1}$ and $B_{1}\bar{B}_{1}$
pairs is suppressed. 
The observed effects of isospin asymmetry of the
medium on the mass modifications of
$D$ and $B$ mesons can be verified
experimentally through the ratios
$\frac{D^{*+}}{D^{*0}}$, $\frac{B^{*+}}{B^{*0}}$,
$\frac{D_{1}^{+}}{D_{1}^{0}}$ and 
$\frac{B_{1}^{+}}{B_{1}^{0}}$, whereas the
effects of strangeness of the matter
can be seen through the 
ratios $\frac{D^{*}}{D_{S}^{*}}$, $\frac{B^{*}}{B_{S}^{*}}$,
$\frac{D_{1}}{D_{1S}}$ and $\frac{B_{1}}{B_{1S}}$.
The traces of observed medium modifications
of masses and decay constants
can be seen experimentally
in the strong decay  width and leptonic decay width of
heavy mesons 
\cite{azzi}. For example, in Ref. \cite{bel1}
the couplings $g_{D^{*}D \pi}$ and
$g_{B^{*}B \pi}$ were studied
using the QCD sum rules and 
strong decay width of charged vector $D^{* +}$ mesons
for the strong decay, $D^{*+}$ $\longrightarrow$ $D^{0} \pi^{+}$
were evaluated using the formula,
\begin{equation}
\Gamma(D^*\longrightarrow D\pi)=\frac{g^2 _{D^*D \pi}}{24 \pi m^2 _{D^*}} |k_{\pi}|^3,
\label{decayw1}
\end{equation}
where, pion momentum, $k_{\pi}$ is, 
\begin{equation}
 k_{\pi} = \sqrt{\frac{(m_{D^*} ^2 - m_{D}^2 + m_{\pi}^2)^2}{(2m_{D^*})^2} - m_\pi ^2}.
 \label{decayw2}
\end{equation}
In Eq. \ref{decayw1}, coupling $g^2 _{D^*D \pi} = 12.5\pm1$,
$m_{D^{*}}$ and $m_D$ denote the masses of
vector and pseudoscalar meson, respectively.
From Eq. \ref{decayw1}, we observe that
the values of decay width depend upon the  masses of 
vector  and pseudoscalar mesons, i.e., $D^{*}$ and $D$, 
respectively.
Using vacuum values for the masses of $D$ mesons, the values of decay width
$\Gamma(D^{*+}$ $\longrightarrow$ $D^{0} \pi^{+})$ 
is observed to be $32\pm5$ keV \cite{bel1}. 
However, as we discussed in our present work, the
 charmed mesons
get modified in the hadronic medium and this must
lead to the medium modification of decay width
of these mesons.
For example, if we 
consider the in-medium masses of vector
mesons from our present work and for
pseudoscalar mesons
we use the in-medium masses from 
Ref. \cite{amavstranged} (in this reference the
mass modifications of pseudoscalar $D$ mesons were
calculated using the chiral SU(4) model in nuclear
and strange hadronic medium), then in nuclear medium $(f_s = 0)$,
at baryon density, $\rho_B$ = $\rho_0$,
the values of decay width,
$\Gamma(D^{*+}$ $\longrightarrow$ $D^{0} \pi^{+})$, 
are observed to be $219$ keV and $31$ keV
at asymmetry parameter $I = 0$ and $0.5$, respectively.
In strange medium ($f_s = 0.5$), the
above values of decay width will change to
$84$ and $25$ keV at $I = 0$ and $0.5$,
respectively. We observe that the decay width of heavy mesons
vary appreciably because of
medium modification of heavy meson masses. In our future work we shall
evaluate in detail the effects of medium modifications
of masses and decay constants of
heavy charmed and bottom mesons on the
above mentioned experimental observables.

\section{Summary}
In short, we computed the 
 mass shift and shift in decay constants of vector
and axial-vector, charm and bottom mesons, in 
asymmetric hadronic matter,  consisting of nucleon and
hyperons, using
phenomenological chiral model and QCD sum rules,
at zero as well as at finite temperature.
For this, first the 
quark and gluon condensates were calculated using
chiral hadronic model and then, using these values
of condensates as input in QCD sum rules, the
 in-medium properties 
of vector and axial-vector mesons
were evaluated. We observed a negative (positive)  shift  in the masses and decay 
constants of
  vector (axial-vector) mesons.
  The magnitude of shift increases with increase in 
  the density of baryonic density of matter.
The properties of mesons are
seen to be sensitive for the isospin asymmetry 
as well as strangeness fraction of the 
medium.  The isospin asymmetry of the
medium causes the mass-splitting 
between isospin doublets.
The presence of hyperons
in addition to nucleons  
cause more decrease (increase) in the masses and decay constant
of vector (axial-vector) mesons.
The finite 
temperature of the medium (at finite baryonic density)
is observed to cause an increase
in the masses and decay constants
 of  heavy mesons.  
The observed effects on the masses and decay 
constants of heavy vector and axial-vector
mesons may be reflected  
 experimentally
in  the production ratio of open charm mesons as well as in their decay width.
The negative mass shift of
charmed vector mesons as observed
in present calculations may cause the
formation of bound states
with the nuclei
as well as the decay of excited charmonium 
states to $D^{*}\bar{D}^{*}$ pairs causing
charmonium suppression.
The present work on the in-medium mesons properties
may be helpful in understanding the
experimental observables
of CBM and PANDA experiments of
FAIR project at GSI Germany.

\acknowledgements 
 The authors gratefully acknowledge the financial support from
the Department of Science and Technology (DST), Government of India for research project under
 fast track scheme for young scientists (SR/FTP/PS-209/2012).

\end{document}